\documentclass[journal]{IEEEtran}
\usepackage{amsmath,amssymb}
\usepackage{url}
\usepackage{soul}
\usepackage{footnote}
\usepackage[para]{footmisc}
\usepackage{graphicx}
\makesavenoteenv{tabular}
\makesavenoteenv{table}
\makesavenoteenv{table*}

\usepackage[dvipsnames]{xcolor}
\usepackage{hyperref}

\newcommand{\blank}{\left<\text{b}\right>}
\newcommand{\sos}{\left<\text{sos}\right>}
\newcommand{\eos}{\left<\text{eos}\right>}

\newcommand{\x}{\mathbf{x}}
\newcommand{\cv}{\mathbf{v}}
\newcommand{\s}{\mathbf{s}}
\newcommand{\C}{\mathcal{C}}
\newcommand{\Ls}{\mathcal{L}}
\newcommand{\A}{\mathcal{A}}

\newcommand{\h}{\mathbf{h}}
\newcommand{\p}{\mathbf{p}}
\newcommand{\q}{\mathbf{q}}
\newcommand{\tout}{\tau}
\newcommand{\EtoE}{E2E}

\begin{document}
\title{End-to-End Speech Recognition: A Survey}
\author{Rohit~Prabhavalkar,~\IEEEmembership{Member,~IEEE,}
        Takaaki Hori,~\IEEEmembership{Senior Member,~IEEE,}
        Tara N. Sainath,~\IEEEmembership{Fellow,~IEEE,\\}
        Ralf Schl\"{u}ter,~\IEEEmembership{Senior Member,~IEEE,}
        and~Shinji Watanabe,~\IEEEmembership{Fellow,~IEEE}%
}

\maketitle

\begin{abstract}
In the last decade of automatic speech recognition (ASR) research, the introduction of deep learning brought considerable reductions in word error rate of more than 50\% relative, compared to modeling without deep learning. In the wake of this transition, a number of all-neural ASR architectures were introduced. These so-called \emph{end-to-end}  (\EtoE) models provide highly integrated, completely neural ASR models, which rely strongly on general machine learning knowledge, learn more consistently from data, while depending less on ASR domain-specific experience. The success and enthusiastic adoption of deep learning  accompanied by more generic model architectures lead to \EtoE\ models now becoming the prominent ASR approach. The goal of this survey is to provide a taxonomy of \EtoE\ ASR models and corresponding improvements, and to discuss their properties and their relation to the classical hidden Markov model (HMM) based ASR architecture. All relevant aspects of \EtoE\ ASR are covered in this work: modeling, training, decoding, and external language model integration, accompanied by discussions of performance and deployment opportunities, as well as an outlook into potential future developments.
\end{abstract}


\IEEEpeerreviewmaketitle

\section{Introduction}
\label{sec:intro}

The classical\footnote{The term "classical" here refers to the former long-term state-of-the-art ASR architecture based on the decomposition into acoustic and language model and with acoustic modeling based on hidden Markov models.} statistical architecture decomposes an ASR
system into four main components: acoustic feature extraction from
speech audio signals, acoustic modeling, language modeling and search
based on Bayes decision rule \cite{Bayes:1763,jelinek1997statistical}. Classical acoustic modeling is based
on hidden Markov models (HMM) to account for speaking rate variation.
Within the classical approach, deep learning has been introduced to
acoustic and language modeling. In acoustic modeling, deep learning
replaced Gaussian mixture distributions (hybrid HMM
\cite{Bourlard+Morgan:1993,Seide+:2011}) or augmented the acoustic
feature set (nonlinear disciminant/tandem approach
\cite{Fontaine+:1997,Hermansky+:2000}). In language modeling, deep
learning replaced count-based approaches
\cite{Nakamura+:1989,Bengio+:2000,Schwenk+Gauvain:2002}. However, when 
introducing deep learning, the classical ASR architecture was not
yet touched. Classical state-of-the-art ASR systems today are
composed of many separate components and knowledge sources, especially
speech signal preprocessing, methods for robustness w.r.t.\ recording
conditions, phoneme inventories and pronunciation lexica, phonetic
clustering, handling of out-of-vocabulary words, various methods for
adaptation/normalization, elaborate training schedules with different
objectives and incl.\ sequence discriminative training, etc.  The
potential of deep learning on the other hand initiated successful
approaches to integrate formerly separate modeling steps, e.g.\
integrating speech signal preprocessing and feature extraction into
acoustic modeling \cite{Tuske+:2014,Sainath+:2015}. 

More consequently, the introduction of deep learning to ASR also initiated 
research to replace classical ASR architectures based on hidden Markov 
models (HMM) with more integrated joint neural network model structures
\cite{graves2006connectionist,graves2012sequence,chorowski2015attention,chan2016listen}.
These ventures might be seen as trading specific speech processing
models for more generic machine learning approaches to
sequence-to-sequence processing, maybe in a similar way as statistical
approaches to natural language processing used to replace more
linguistically oriented models. For these all-neural approaches
recently the term \emph{end-to-end} (\EtoE)
\cite{Liang+:2006,Collobert+:2011,graves2012sequence,GravesJaitly14} 
has been established. However, it lacks distinction in many ways. 
Therefore, first of all an attempt to defining the term 
\emph{end-to-end} in the context of ASR is due in this survey.

According to the Cambridge Dictionary, the adjective ``end-to-end'' is
defined by: ``including all the stages of a process''
\cite{CambridgeDictEndToEnd}. This can be regarded from a number of perspectives: 
\paragraph{Joint Modeling} In terms of ASR, the \EtoE\ property might be understood as considering all
components of an ASR system jointly as a single computational graph. Even more so, the common
understanding of \EtoE\ in ASR is that of a single joint modeling
approach that does not necessarily distinguish separate components,
which also may mean dropping the classical separation of ASR into an
acoustic model and a language model.

\paragraph{Single-Pass Search} In terms of the recognition/search problem, the \EtoE\ property can be
interpreted as integrating all components (models, knowledge sources)
of an ASR system before coming to a decision. This is in line with
Bayes' decision rule, which exactly requires a single global decision
integrating all available knowledge sources.

\paragraph{Joint Training} In terms of model training, \EtoE\ suggests estimating all parameters
of all components of a model jointly using a single objective
function that is consistent with the task at hand, which in case of
ASR means minimizing the expected word error rate.
\paragraph{Training Data} Joint training of an integrated model implies using a single kind of
training data, which in case of ASR would be transcribed speech audio
data. However, in ASR often even larger amounts of text-only data, as
well as optional untranscribed speech audio are available. One of the
challenges of \EtoE\ modeling therefore is how to take advantage of
text-only and audio-only data \cite{tjandra2017speechchain,baskar2019semisupervised}.

\paragraph{Training from Scratch} The \EtoE\ property can also be interpreted for the training process
itself, by requiring training \emph{from scratch} avoiding external
knowledge like prior alignments or initial models pre-trained using
different criteria and/or knowledge sources.
Note that pre-training and fine-tuning strategies are also important in some scenarios, if the model has explicit modularity, including self-supervised learning~\cite{Baevski2020wav2vec2} or joint training of front-end and speech recognition models~\cite{chang2022end}.

\paragraph{Secondary Knowledge Sources} For ASR, standard secondary knowledge sources are pronunciation lexica
and phoneme sets, as well as phonetic clustering, which in classical
state-of-the-art ASR systems usually is based on classification and
regression trees (CART) \cite{Breiman}. Secondary knowledge sources
and separately trained components may introduce errors, might be
inconsistent with the overall training objective and/or may generate
additional cost. Therefore, in an \EtoE\ approach, these would be
avoided.

\paragraph{Vocabulary Modeling} Avoiding pronunciation lexica and corresponding subword units would
limit \EtoE\ recognition vocabularies to be based on whole word or
character models. Whole word models \cite{soltau2016neural}, according
to Zipf's law \cite{Zipf}, would require unrealisticly high amounts of
transcribed training data for large vocabularies, which might not be
attainable for many tasks. On the other hand, methods to generate
subword vocabularies based on characters, like the currently popular
byte pair encoding (BPE) approach \cite{Sennrich+:2015}, might be seen
as secondary approaches outside the \EtoE\ objective.

\paragraph{Generic vs.\ Informed Modeling} Finally, \EtoE\ may also be seen in terms of the genericity of the
underlying modeling: are task-specific constraints learned from data
completely, or does task-specific knowledge enter modeling the system
architecture in the first place? For example, the monotonicity
constraint in ASR may be learned completely from data like in
attention-based \EtoE\ approaches \cite{chan2016listen}, or it may
directly be implemented, as in classical HMM structures.

Overall, end-to-end ASR therefore can be defined as an
\emph{integrated ASR model that enables joint training and recognition
  consistently minimizing expected word error rate, avoiding
  separately obtained knowledge sources}.

However, what are potential benefits of \EtoE\ approaches to ASR? The
primary objective when developing ASR systems is to minimize the
expected word error rate. However, secondary objectives are to reduce
time and memory complexity of the resulting decoder, and - assuming a
constrained development budget - genericity and ease of modeling.

First of all, defining an ASR system \EtoE\ in terms of a single
neural network structure supports modeling genericity and may allow
for faster development cycles when building ASR systems for new
languages/domains. ASR models defined by a single neural network
structure may become more lean compared to classical modeling, and
the decoding process becomes simpler as it does not need to integrate
separate models. The resulting reduction in memory footprint and power
consumption supports embedded ASR applications \cite{Pang+:2018}.

Furthermore, joint training \EtoE\ may help to avoid spurious optima from
intermediate training stages. Avoiding secondary knowledge sources
like pronunciation lexica may be helpful for languages/domains where
such resources are not easily available. Also, secondary knowledge
sources may itself be erroneous. Avoiding these supports improved
models trained directly from data, provided that sufficient amounts of
task-specific training data are available.

With the current surge of interest in \EtoE\ ASR models and an increasing diversity of corresponding work, the authors of this review think it is time to provide an overview of this rapidly evolving domain of research.
The goal of this survey is to provide an in-depth overview of the current state of research w.r.t.\ \EtoE\ ASR systems, covering all relevant aspects of \EtoE\ ASR and including a contrastive discussion of the different \EtoE\ and classical ASR architectures.

This survey of \EtoE\ speech recognition is structured as
follows. Sec.~\ref{sec:e2e}
describes the historical evolution of \EtoE\ speech recognition, with
specific focus on the input/output alignment and an overview of
currently prominent basic \EtoE\ ASR models.  
Sec.~\ref{sec:improvements}
discusses improvements of the basic \EtoE\ models, incl.\ \EtoE\ model
combination, training loss functions, context, encoder/decoder
structures and endpointing.  
Sec.~\ref{sec:training}
provides an overview of \EtoE\ ASR model training.
Decoding algorithms for the different \EtoE\ approaches are
discussed in Sec.~\ref{sec:decoding}.
Sec.~\ref{sec:lm}
discusses the role and integration of (separate) language models in \EtoE\
ASR.  
In Sec.~\ref{sec:relationship},
the relationship between novel \EtoE\ models and classical,
HMM-based ASR approaches are discussed.
Sec.~\ref{sec:comparison}
reviews experimental comparisons of the different \EtoE\ as well as
classical ASR approaches.
Sec.~\ref{sec:applications}
gives an overview of applications of \EtoE\ ASR approaches. Sec.~\ref{sec:future}
investigates future directions of \EtoE\ research in ASR, followed by final conclusions in
Sec.~\ref{sec:conclusions}.

\section{A Taxonomy of E2E Models in ASR}
\label{sec:e2e}

Before we discuss the details of various E2E modeling approaches, we first introduce our notation.
We denote the input speech utterance as $X$, which we assume has been parameterized into $D$-dimensional acoustic frames (e.g., log-mel features) of length $T'$: $X = (\x_1, \cdots, \x_{T'})$, where $\x_t \in \mathbb{R}^{D}$.
We denote the corresponding word sequences as $C$, which can be decomposed into a suitable sequence of labels of length $L$: $C = (c_1, \cdots, c_L)$, where each label $c_j \in \C$.
Our description is agnostic to the specific representation used for decomposing the word sequence into labels; popular choices include characters, words, or sub-word sequences (e.g., BPE~\cite{Sennrich+:2015}, word-pieces~\cite{SchusterNakajima02}).
All E2E models consist of an \emph{encoder} module, denoted by $H(X)$, which maps the input acoustic frame sequence of length $T'$ into a higher-level representation, $H(X) = (\h_1, \cdots, \h_T)$ of length $T$ (typically $T \leq T'$).
Once again, our discussion is agnostic to the specific modeling choices that are involved in the design of the neural network used to implement the function $H(X)$ -- e.g., this might involve recurrent neural networks (either uni-directional or bi-directional)~\cite{hochreiter1997long}, convolutional networks, or transformers~\cite{vaswani2017attention}. 
Given a set of training examples, $\mathcal{T}$, consisting of $N$ pairs of utterances and corresponding transcripts: $\mathcal{T} = \{(X_i, C_i)\}_{i=1}^N$, our goal is to train a model to estimate the conditional distribution over all possible transcripts, given the input acoustics: $P(C|X) = P(C | H(X))$.

Broadly speaking, one of the challenges in estimating $P(C|X)$ with E2E models is the need to account for the unknown alignments between the acoustic frame sequences and the corresponding label sequences of length $T$ and $L$, respectively.
In classical ASR models, these frame-level alignments can be modeled with hidden Markov models (HMMs), while using generative Gaussian mixture models (GMMs) or neural networks to model the output distribution of acoustic frames; frame-level alignments to train neural network acoustic models may be obtained by force-alignment from a base GMM-HMM systems, but direct sequence training not requiring initial alignments also is possible~\cite{Povey+Peddinti+:2016}.
While alignment for phonetic label sequences, as usually done in classical ASR systems, seems natural; E2E-based systems usually employ (sub-)word units based on characters. However, a notion of character-level alignments seems less clear\footnote{For example, in languages such as English the correspondence between orthography and pronunciation is not straightforward: a word such as \texttt{knife} can be decomposed into it's constituent phonemes \texttt{n ay f} which can be associated to individual acoustic frames, but it is less clear what frames should correspond to the silent character \text{k} in the word when considering it's representation as a character sequence.}. Therefore, in this work, we categorize various E2E modeling approaches based on how the system models this alignment -- whether this is done explicitly or implicitly. 

Early E2E modeling approaches \emph{modeled alignments explicitly} through a latent variable, which is marginalized out (possibly, approximately) during training and inference.
Examples of this family of approaches include connectionist temporal classification (CTC)~\cite{graves2006connectionist}, the recurrent neural network transducer (RNN-T)~\cite{graves2012sequence}, the recurrent neural aligner (RNA)~\cite{sak2017recurrent}, and the hybrid auto-regressive transducer~\cite{VarianiRybachAllauzen+20} (HAT).
As will be discussed in subsequent sections, the latter modeling approaches in this family represent increasingly sophisticated modeling of alignments, with fewer independence assumptions and are thus increasingly powerful.
At the other end of the spectrum, we find attention-based encoder-decoder models which were first popularized in the context of machine translation~\cite{WuSchusterChen+16}. 
Unlike explicit alignment models such as CTC, attention-based encoder-decoder models use an attention mechanism~\cite{BahdanauChoBengio14} to learn a correspondence between the entire acoustic sequence and the individual labels.
Finally, there is a body of work that lies in between these two extremes: models such as the neural transducer~\cite{JaitlyLeVinyal+s16}, or those based on monotonic alignments~\cite{raffel2017online} and its variants (e.g., monotonic chunkwise alignments (MoChA)~\cite{chiu2018monotonic}, monotonic infinite lookback (MILK)~\cite{arivazhagan2019monotonic} etc.) use an explicit alignment model, while also utilizing an attention mechanism that allows the model to \emph{examine local acoustics} in order to refine predictions.
We describe each of these models in order in the remainder of this section.

\subsection{Explicit Alignment E2E Approaches}
\label{sec:explicit-alignment-e2e-approaches}
Explicit alignment approaches such as CTC, RNN-T, or RNA, define an explicit latent variable corresponding to the alignments between the $T$-length encoder output, $H(X)$, and the $L$-length output sequence; each approach differs in how it defines the alignment which in turn determines the underlying probabilistic conditional independence assumptions and the process of training and decoding in the respective models.
A common feature of all explicit alignment models is that they introduce an additional \emph{blank} symbol, denoted $\blank$, and define an output probability distribution over symbols in the set $\C_b = \C \cup \left\{ \blank\right\} $.
The interpretation of the $\blank$ symbol varies slightly between each of these models, as we discuss in greater details below. 
For now, it suffices to say that given a specific training example, $(X, C)$, each of these models defines a set of \emph{valid alignments}, denoted by $\A_{(T, C)}$, and defines the conditional distribution $P(C|X)$ by marginalizing over all valid alignment sequences:
\begin{align}
    P(C|X) = P(C|H(X)) &= \sum_A P(C | A, H(X)) P(A|H(X)) \nonumber \\
    &= \sum_{A \in \A_{(T=|H(X)|, C)}} P(A | H(X)) \label{eq:explicit-alignment-approaches}
\end{align}
\noindent where, by definition $P(C | A, H(X)) = 1$ if and only if $A \in \A_{(T, C)}$ and $0$ otherwise, which also means that the mapping from an alignment $A$ to a label sequence $C$ is defined to be unique.
We discuss the specific formulations of each of these models in the subsequent sections.

\subsubsection{Connectionist Temporal Classification (CTC)}
\label{sec:ctc}
\begin{figure}
    \centering
    \includegraphics[width=0.45\textwidth]{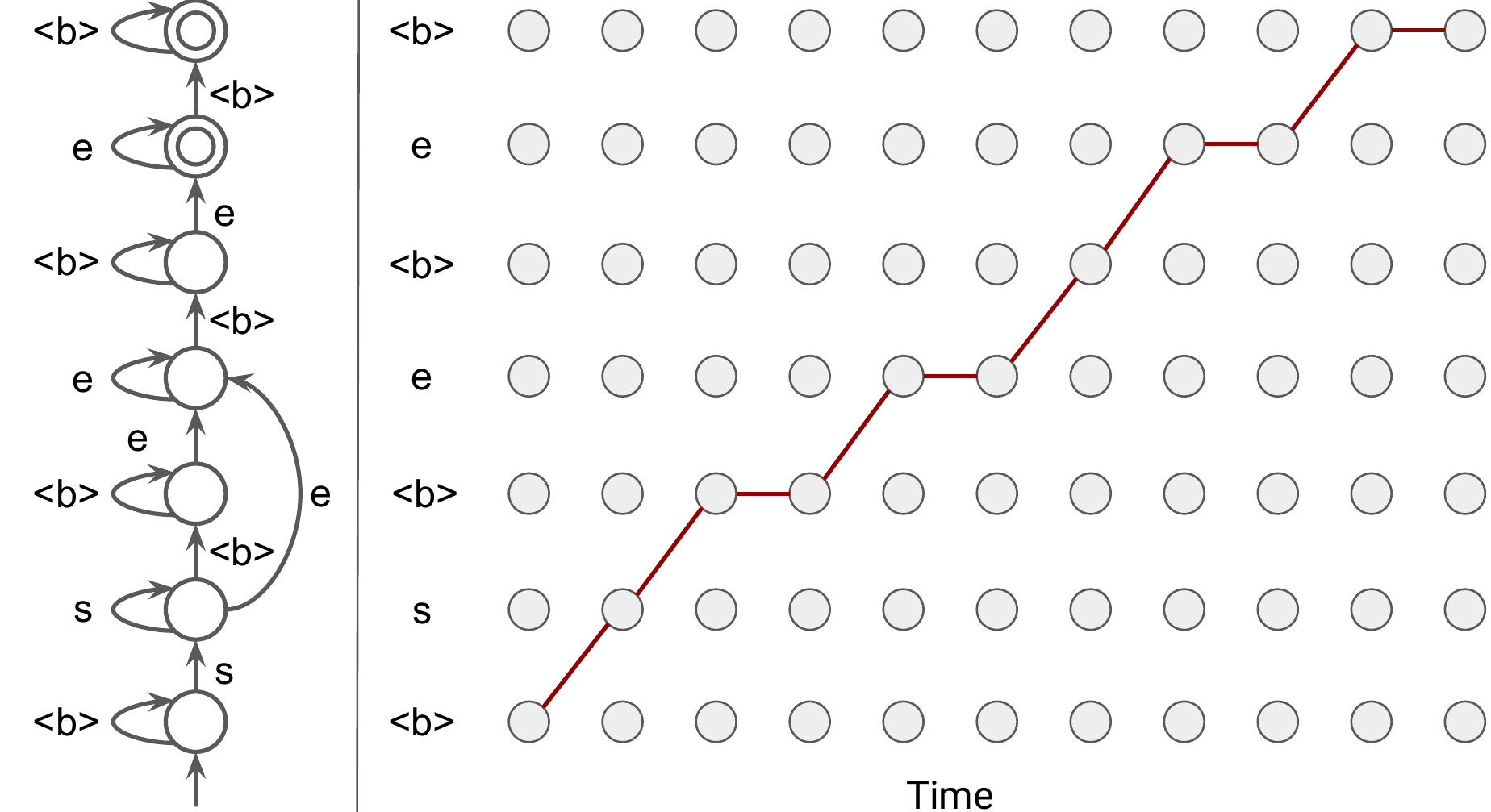}
    \caption{Example alignment sequence for a CTC model with the target sequence $C = (\texttt{s}, \texttt{e}, \texttt{e})$ (right), alongside a (non-deterministic) finite state automaton (FSA)~\cite{mohri2002weighted} (left) representing the set of all valid alignment paths.}
    \label{fig:ctc-alignment}
\end{figure}

Connectionist Temporal Classification (CTC) was proposed by Graves et al.~\cite{graves2006connectionist} as a technique for mapping a sequence of input tokens to a corresponding sequence of output tokens.
CTC explicitly models alignments between the encoder output, $H(X)$, and the label sequence, $C$, by introducing a special ``blank" label, denoted by $\blank$: $\C_b = \C \cup \left\{ \blank\right\} $.
An alignment, $A \in \C_b^*$, is thus a sequence of labels in $\C$ or $\blank$. 
Given a specific training example, $(X, C)$, we denote the set of all valid alignments, $\mathcal{A}^{\text{CTC}}_{(X, C)} = \{A = (a_1, a_2, \ldots, a_T)\}$, such that each $a_t \in \C_b$ with the additional constraint that $A$ is identical to $C$ after first collapsing consecutive identical labels, and then removing all blank symbols.
For example, if $T=10$, and $C = (\texttt{s}, \texttt{e}, \texttt{e})$, then $A = (\texttt{s}, \blank, \blank, \texttt{e}, \texttt{e}, \blank, \texttt{e}, \texttt{e}, \blank, \blank) \in \mathcal{A}^{\text{CTC}}_{(X, C)}$, as illustrated in Figure~\ref{fig:ctc-alignment}.
As can be seen in this example, repeated labels in the output can be represented by intervening blanks.

Following Eq.~\eqref{eq:explicit-alignment-approaches}, CTC defines the posterior probability of the label sequence $C$ conditioned on the input, $X$, by marginalizing over all possible CTC alignments as:
\begin{align}
    P_\text{CTC}(C|X) 
        &= \sum_{A \in \mathcal{A}^{\text{CTC}}_{(X, C)}} P(A|H(X)) \nonumber \\
        &= \sum_{A \in \mathcal{A}^{\text{CTC}}_{(X, C)}} \prod_{t=1}^T P(a_t | a_{t-1}, \cdots, a_1, H(X)) \nonumber \\
        &= \sum_{A \in \mathcal{A}^{\text{CTC}}_{(X, C)}} \prod_{t=1}^T P(a_t | \h_t) \label{eq:ctc-independence-assumption}
\end{align}
\noindent Critically, as can be seen in Eq.~\eqref{eq:ctc-independence-assumption}, CTC makes a strong independence assumption that the model's output at time $t$ is conditionally independent of the outputs at other timesteps, given the local encoder output at time $t$.

\begin{figure}
     \centering
     \includegraphics[width=0.2\textwidth]{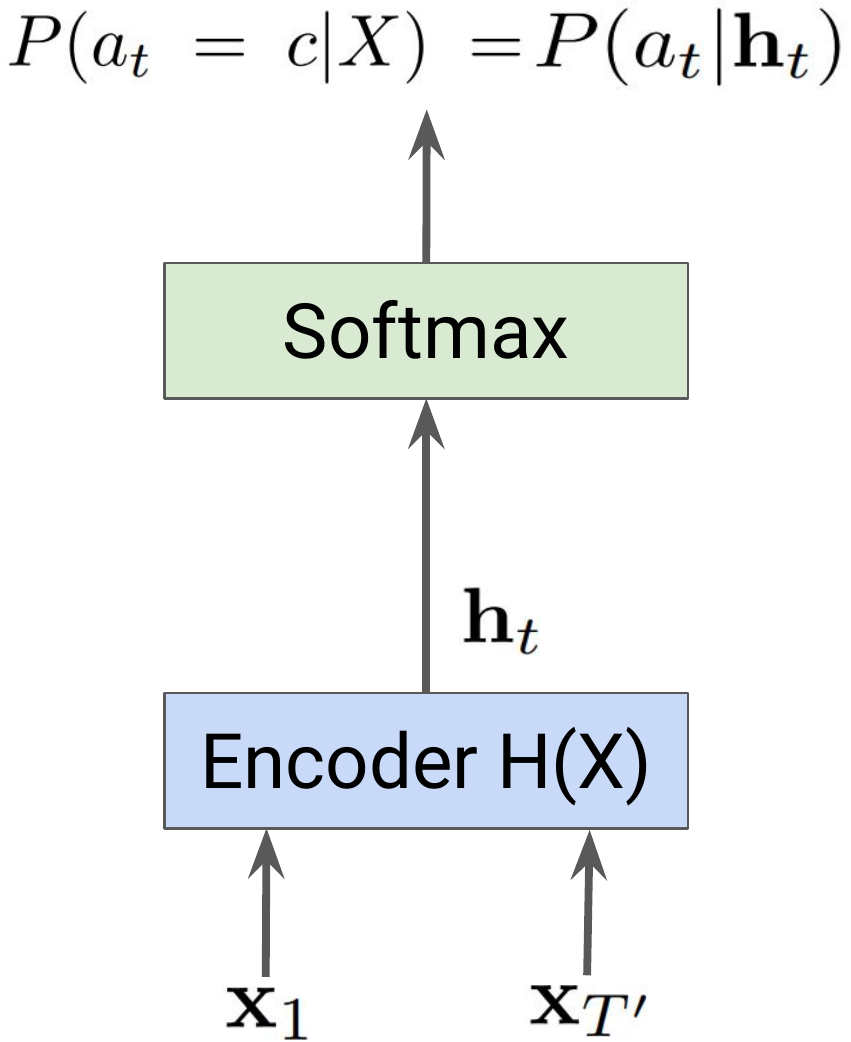}
     \caption{A representation of the CTC model consisting of an encoder which maps the input speech into a higher-level representation, and a softmax layer which predicts frame-level probabilities over the set of output labels and blank.}
     \label{fig:ctc}
\end{figure}
Thus, a CTC model consists of a neural network that models the distribution $P(a_t | X)$, at each step as shown in Figure~\ref{fig:ctc}. 
For example, this can be accomplished using a neural network, which we refer to as the encoder; popular choices include deep convolutional networks, or deep recurrent neural network (RNN) such as a network with long short-term memory (LSTM) cells~\cite{hochreiter1997long}.
The encoder (which may be uni- or bi-directional) converts the input acoustic frames, X, into an encoded representation, $H = (\h_1, \cdots, \h_T)$, where $h_t \in \mathbb{R}^{k}$.
The encoder is connected to a softmax layer with $|\C_b|$ targets representing the individual probabilities in Eq.~\eqref{eq:ctc-independence-assumption}: $P(a_t = c | X) = P(a_t = c | H(X))$.
Thus, at each step, $t$, the model consumes a single encoded frame $h_t$ and outputs a distribution over the labels; in other words, the model ``outputs" a single label either blank, $\blank$, or one of the targets in $\C$.

\subsubsection{Recurrent Neural Network Transducer (RNN-T)}
\begin{figure}
     \centering
     \includegraphics[width=0.27\textwidth]{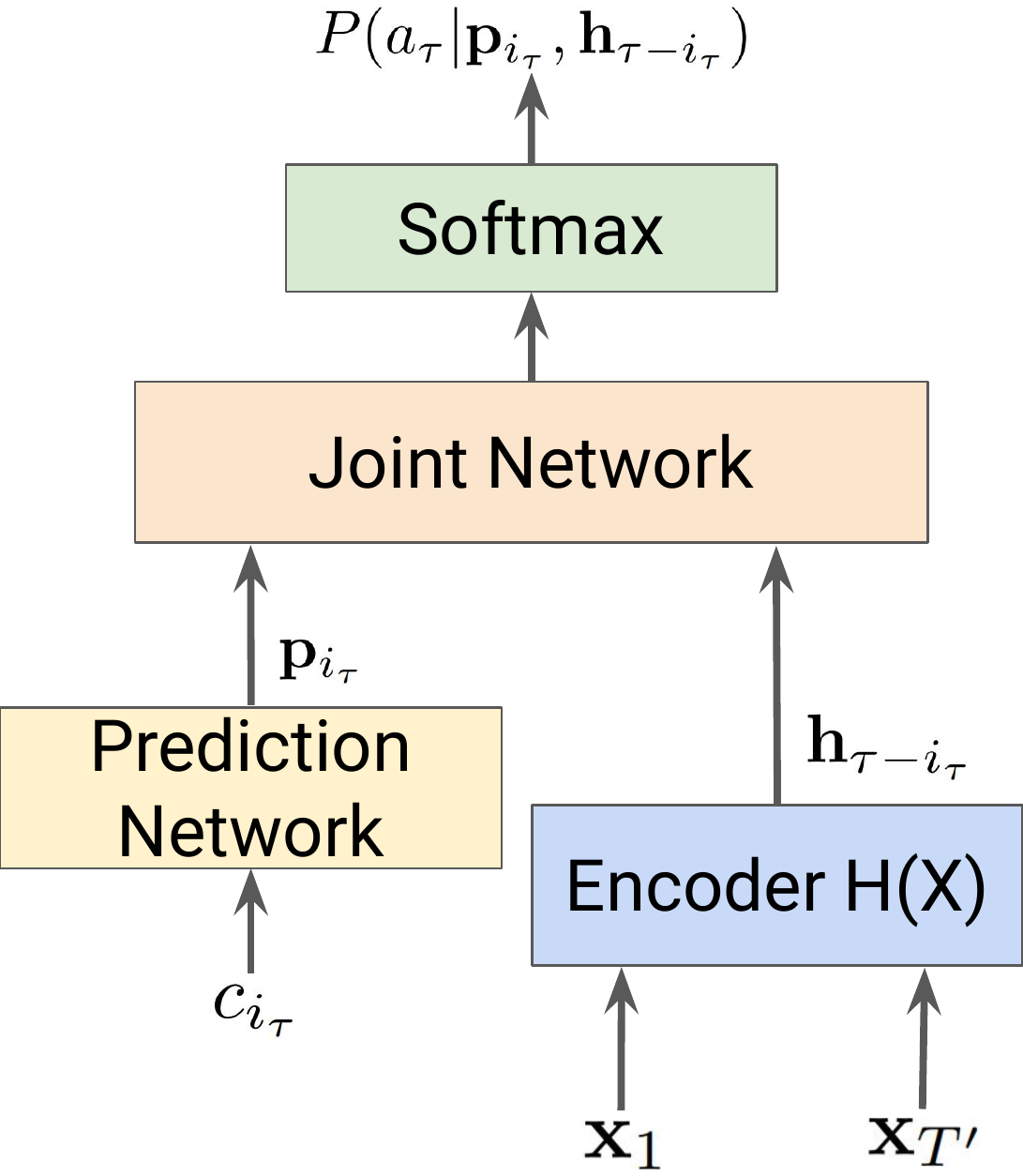}
     \caption{An RNN-T Model~\cite{graves2012sequence, graves2013speech} consists of an encoder which transforms the input speech frames into a high-level representation, and a prediction-network which models the sequence of non-blank labels that have been output previously. The prediction network output, $p_{i_t}$, represents the output after producing the previous non-blank label sequence $c_1, \ldots, c_{i_t}$. The joint network produces a probability distribution over the output symbols (augmented with blank) given the prediction network state and a specific encoded frame.}
     \label{fig:rnnt}
\end{figure}
The Recurrent Neural Network Transducer (RNN-T)~\cite{graves2012sequence, graves2013speech} was proposed by Graves as an improvement over the basic CTC model~\cite{graves2006connectionist}, by removing some of the conditional independence assumptions that we discussed previously.
The RNN-T model, which is depicted in Figure~\ref{fig:rnnt}, is best understood by contrasting it against the CTC model.
As with CTC, the RNN-T model augments the output symbols with the blank symbol, and thus defines a distribution over label sequences in $\C_b$.
Similarly, as with CTC, the model consists of an encoder which processes the input acoustic frames X to generate the encoded representation $H(X) = (\h_1, \cdots, \h_T)$.

Unlike CTC, however, the blank symbol in RNN-T has a slightly different interpretation; for each input encoder frame, $\h_t$, the RNN-T model outputs a sequence of zero or more symbols in $\C$ which are terminated by a single blank symbol. 
Thus, we may define the set of all valid alignment sequences in RNN-T as: $\mathcal{A}^{\text{RNNT}}(X, C) = \{A = (a_1, a_2, \cdots, a_{T+L}) \}$, the set of all sequences of $T + L$ symbols in $\C_b^*$, which are identical to $C$ after removing all blanks.
Finally, for a given output position $\tout$, let $i_{\tout}$ denote the number of non-blank labels in the partial sequence $(a_1, \cdots, a_{\tout-1})$.
Thus, the number of blanks in the partial sequence $(a_1, \cdots, a_{\tout-1})$ is $\tout -i_{\tout} - 1$.
For example, if $T=7$, and $C=(\texttt{s}, \texttt{e}, \texttt{e})$, then $A = (\blank, \texttt{s}, \blank, \blank, \blank, \texttt{e}, \texttt{e}, \blank, \blank, \blank) \in \mathcal{A}^{\text{RNNT}}_{(X, C)}$.
\begin{figure}
    \centering
    \includegraphics[width=0.45\textwidth]{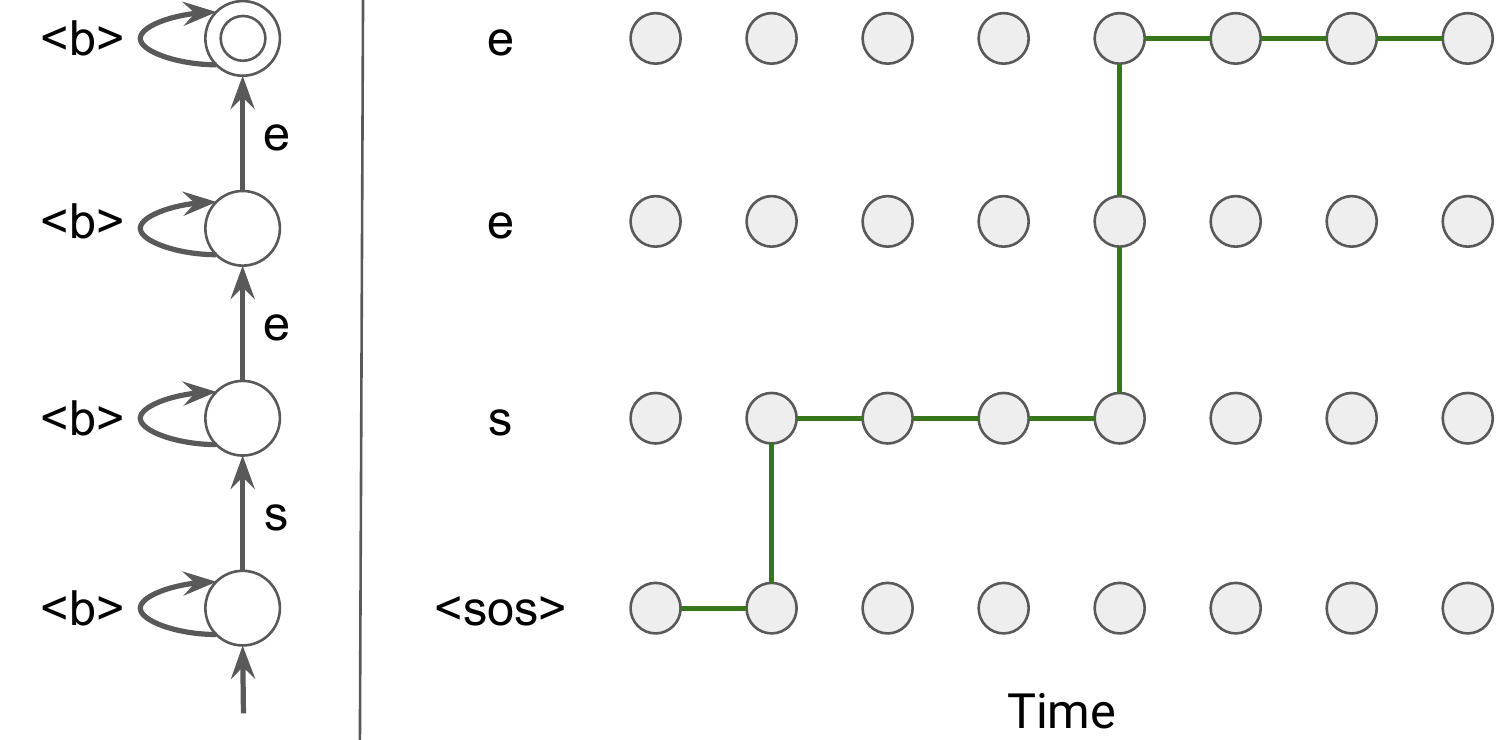}
    \caption{Example alignment sequence (right) for an RNN-T model with the target sequence $C = (\texttt{s}, \texttt{e}, \texttt{e})$. Horizontal transitions in the image correspond to blank outputs. The FSA (left) represents the set of all valid RNN-T alignment paths.}
    \label{fig:rnnt-alignment}
\end{figure}
Note that, unlike the CTC model, repeated labels in the output require no special treatment as illustrated in Figure~\ref{fig:rnnt-alignment}; $i_1 = i_2 = 0; i_3 = i_4 = 1; i_{10} = 3;$ etc.

We may then define the posterior probability $P(C|X)$ as before:
\begin{align}
P_\text{RNNT}(C|X) &= \sum_{A \in \mathcal{A}^{\text{RNNT}}_{(X, C)}} P(A|H(X)) \nonumber \\
    &= \sum_{A \in \mathcal{A}^{\text{RNNT}}_{(X, C)}} \prod_{\tout=1}^{T+L} P(a_{\tout} | a_{\tout-1}, \ldots, a_1, H(X)) \nonumber \\
    &= \sum_{A \in \mathcal{A}^{\text{RNNT}}_{(X, C)}} \prod_{\tout=1}^{T+L} P(a_{\tout} | c_{i_{\tout}}, c_{i_{\tout} - 1}, \ldots, c_0, \h_{\tout - i_{\tout}}) \label{eq:rnnt-independence-assumption} \\
    &= \sum_{A \in \mathcal{A}^{\text{RNNT}}_{(X, C)}} \prod_{\tout=1}^{T+L} P(a_{\tout} | \p_{i_{\tout}}, \h_{\tout - i_{\tout}}) \nonumber 
\end{align}
\noindent 
where, $P = (\p_1, \cdots, \p_L)$ represents the output of the \emph{prediction network} depicted in Figure~\ref{fig:rnnt} which summarizes the sequence of previously predicted non-blank labels, implemented as another neural network: $\p_j = NN(\cdot|c_0, \ldots, c_{j-1})$, where $c_0$ is a special start-of-sentence label, $\sos$.
Thus, as can be seen in Eq.~\eqref{eq:ctc-independence-assumption}, RNN-T reduces some of the independence assumptions in CTC since we model the output at time t is conditionally dependent on the sequence of previous non-blank predictions, but is independent of the specific choice of alignment (i.e., the choice of the frames at which the non-blank tokens were emitted).

Finally, we note that the work in~\cite{Moritz+Hori+:2022} generalizes the basic RNN-T model by allowing the set of valid frame-level alignments to be represented as an arbitrary graph; the forward-backward algorithm required to compute the conditional probability distributions can be generalized suitably over the graph.
This formulation, for example, allows for the use of ``ctc-like" alignments in the RNN-T model (i.e., outputting a single label -- blank, or non-blank -- at each frame) while conditioning on the set of previous non-blank symbols as in the RNN-T model.

\subsubsection{Recurrent Neural Aligner (RNA)}

\begin{figure}
     \centering
     \includegraphics[width=0.27\textwidth]{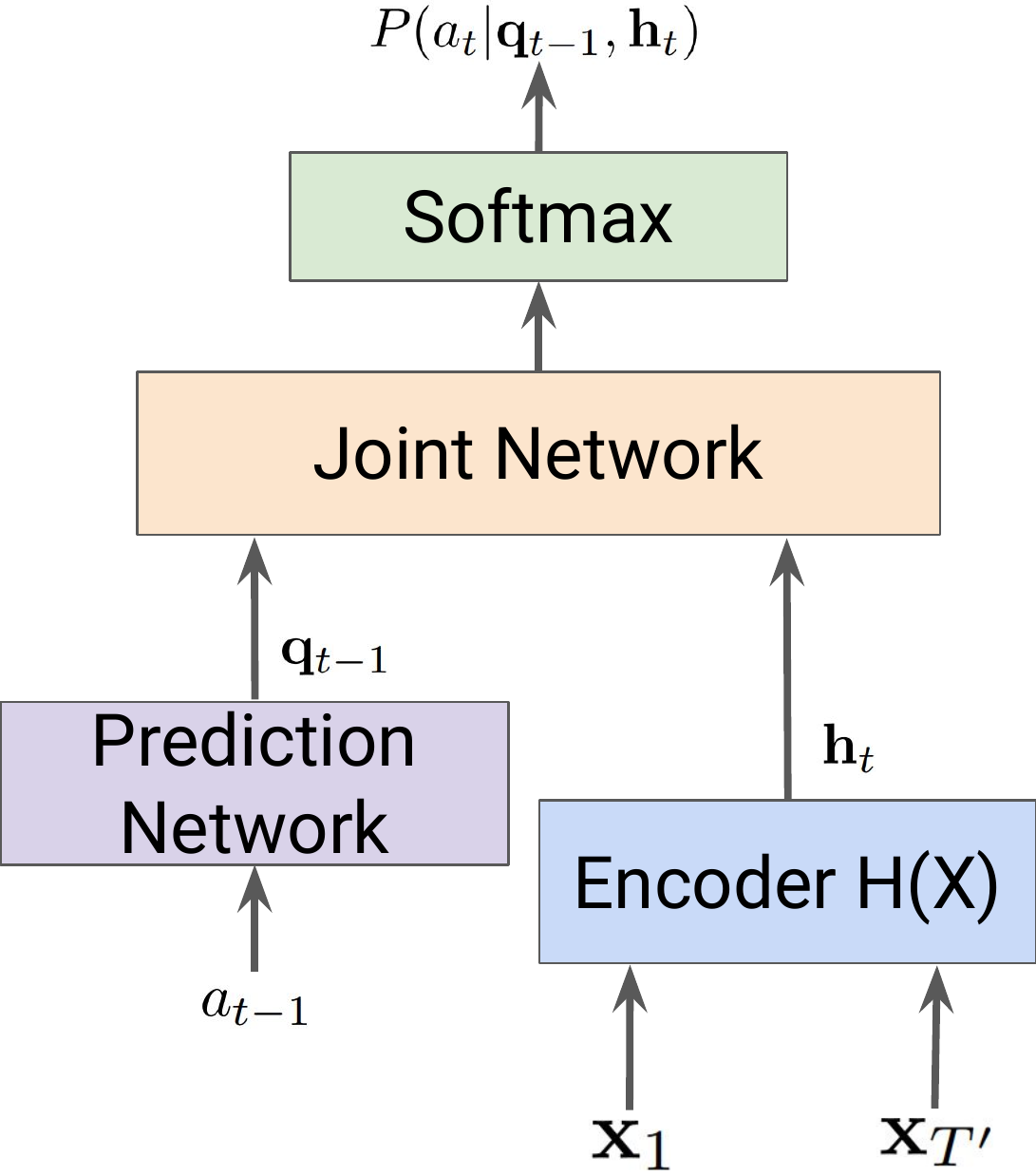}
     \caption{An RNA Model~\cite{sak2017recurrent} resembles the RNN-T model~\cite{graves2012sequence, graves2013speech} in terms of the model structure. However, this model is only permitted to output a single label -- either blank, or non-blank -- in a single frame. Unlike RNN-T, the prediction network state in the RNA model, $\q_{t-1}$, depends on the entire alignment sequence $a_{t-1}, \ldots, a_1$. The joint network produces a probability distribution over the output symbols (augmented with blank) given the prediction network state and a specific encoded frame.}
     \label{fig:rna}
\end{figure}
The recurrent neural aligner (RNA) was proposed by Sak et al.~\cite{sak2017recurrent}.
The RNA model generalizes the RNN-T model by removing one of its conditional independence assumptions.
The model, depicted in Figure~\ref{fig:rna}, is best understood by considering how it differs from the RNN-T model.
As with CTC and RNN-T, the RNA model defines a probability distribution over blank augmented labels in the set $\C_b$, where $\blank$ has the same semantics as in the CTC model: at each frame the model can only output a single label -- either blank, or non-blank -- before advancing to the next frame; unlike CTC (but as in RNN-T) the model only outputs a single instance of each non-blank label.
More specifically, the set of valid alignments, $\A^\text{RNA}_{(X, C)} = (a_1, \cdots, a_{T})$, in the RNA model consist of length $T$ sequences in $\C_{b}^{*}$ with exactly $T-L$ blank symbols, and which are identical to $C$ after removing all blanks.
Thus, the blank symbol has a different interpretation in RNA and the RNN-T models: in RNN-T, outputting a blank symbol advances the model to the next frame; in RNA, however, the model advances to the next frame after outputting a single blank or non-blank label.
Restricting the model to output a single non-blank label at each frame improves computational efficiency and simplifies the decoding process, by limiting the number of model expansions at each frame (cf. RNN-T decoding). 
\begin{figure}
    \centering
    \includegraphics[width=0.45\textwidth]{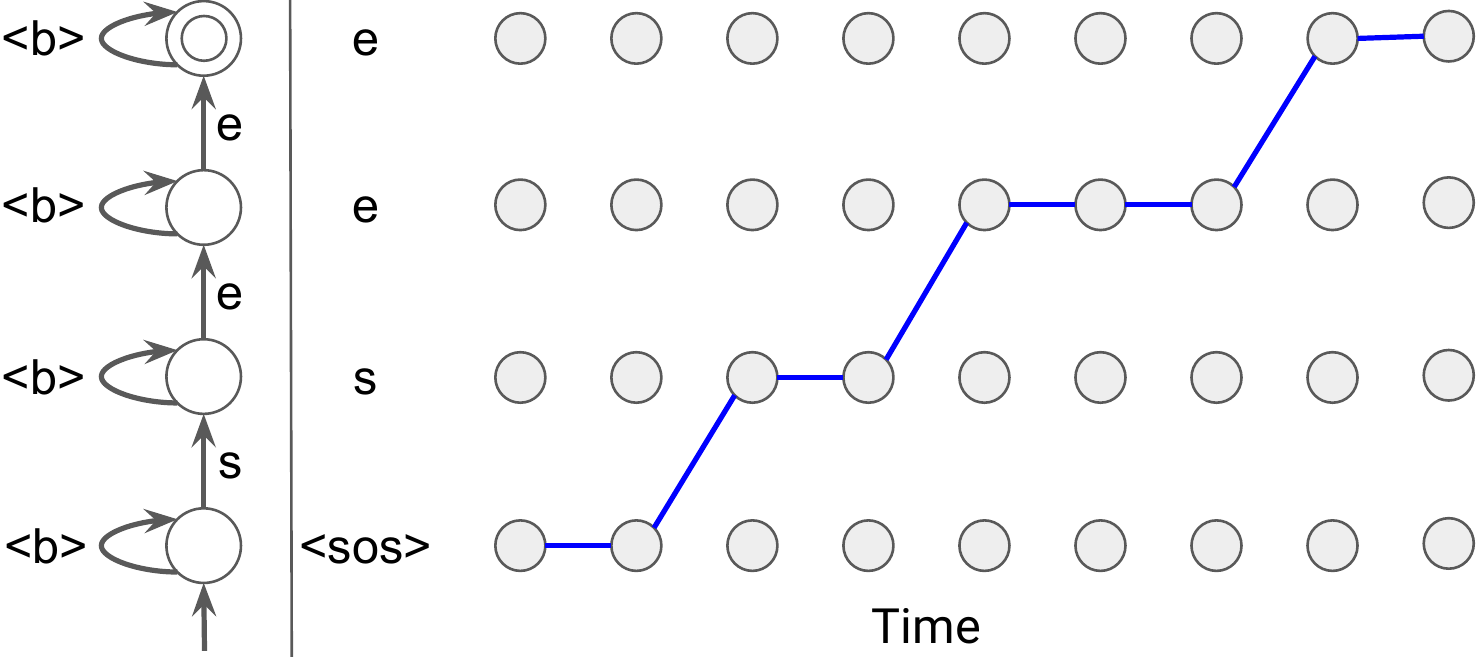}
    \caption{Example alignment sequence (right) for an RNA model with the target sequence $C = (\texttt{s}, \texttt{e}, \texttt{e})$. Horizontal transitions in the image correspond to blank outputs; diagonal transitions correspond to outputting a non-blank symbol. The FSA (left) represents the set of valid alignments for the RNA model. Although the FSA is identical to the corresponding FSA for RNN-T in Figure~\ref{fig:rnnt-alignment}, the semantics of the $\blank$ label are different in the two cases.}
    \label{fig:rna-alignment}
\end{figure}
For example, if $T=8$, and $C=(\texttt{s}, \texttt{e}, \texttt{e})$, then $A = (\blank, \texttt{s}, \blank, \texttt{e}, \blank, \blank, \texttt{e}, \blank) \in \mathcal{A}^{\text{RNA}}_{(X, C)}$ as illustrated in Figure~\ref{fig:rna-alignment}.

The RNA posterior probability, $P(C|X)$, is defined as:
\begin{align}
P_\text{RNA}(C|X) &= \sum_{A \in \mathcal{A}^{\text{RNA}}_{(X, C)}} P(A|H(X)) \nonumber \\
    &= \sum_{A \in \mathcal{A}^{\text{RNA}}_{(X, C)}} \prod_{t=1}^{T} P(a_t | a_{t-1}, \ldots, a_1, H(X)) \nonumber \\
    &= \sum_{A \in \mathcal{A}^{\text{RNA}}_{(X, C)}} \prod_{t=1}^{T} P(a_t | \q_{t-1}, \h_{t}) \label{eq:rna-independence-assumption}
\end{align}

\noindent where, as before $i_t$ denotes the number of non-blank symbols in the partial alignment sequence $(a_1, \ldots, a_{t-1})$, and $\q_{t-1} = \text{NN}(\cdot|a_{t-1}, \cdots, a_1)$ represents the output of a neural network which summarizes the entire partial alignment sequence, where $\text{NN}(\cdot)$ represents a suitable neural network (an LSTM in~\cite{sak2017recurrent}).
Thus, RNA removes the one remaining conditional independence assumption of the RNN-T model, by conditioning on the sequence of previous non-blank labels \emph{as well as the alignment that generated them}.

Since the RNA model conditions on both the sequence of non-blank labels (as in RNN-T), as well as the specific frames at which these are emitted, the exact computation of the log-likelihood in Eq.~\eqref{eq:rnnt-independence-assumption} (and corresponding gradients) is intractable.
Instead, RNA makes a simplifying assumption by utilizing the RNN-T states in the decoder RNN $\text{NN}(\cdot)$ corresponding to the most likely alignment path, while exploiting the constraint that the model can only output a single label at each frame.
The latter constraint -- allowing only a single label (blank or non-blank) at each frame -- has also been explored in the context of the monotonic RNN-T model~\cite{tripathi2019monotonic}.
Finally, we note that the work in~\cite{zeyer2020new} further generalizes the RNA model by employing two RNNs when defining the state: a slow RNN (which corresponds to the sequence of previously predicted non-blank labels), and a fast RNN (which also conditions on the frames at which the non-blank labels were output).

\subsection{Implicit Alignment E2E Approaches}
One of the main benefits of the explicit alignment approaches such as CTC, RNN-T, or RNA is that they result in ASR models that are easily amenable to \emph{streaming} -- i.e., assuming that the neural network used to generate encoder features $H(X)$ can produce encoded frames based on only the sequence of previous acoustic features with limited look-ahead, then the models can output hypotheses as speech is input to the system.
However, there are a number of applications where \emph{offline speech processing} may be appropriate: i.e., where all of the input speech is fed into the system before any outputs are produced.
In this section, we discuss the attention-based encoder-decoder (AED) models (also known as, listen-attend-and-spell (LAS))~\cite{chorowski2015attention, chan2016listen, bahdanau2016end}, which employs the \emph{attention mechanism}~\cite{BahdanauChoBengio14} to implicitly identify and model the portions of the input acoustics which are relevant to each output unit.
The attention mechanism thus avoids the need for explicit modeling of alignments in the basic AED model.

In the streaming explicit alignment approaches presented in Section~\ref{sec:explicit-alignment-e2e-approaches}, during inference, the model continues to output symbols until it has processed the final frame at which point the decoding process is complete; similarly, during training, the forward-backward algorithm aligns over all possible alignment sequences.
Since an AED model processes the entire acoustic sequence at once, the model needs a mechanism by which it can indicate that it is done emitting all output symbols. 
This is achieved by augmenting the set of outputs with an end-of-sentence symbol, $\eos$, so that the output vocabulary consists of the set $\C_{\text{eos}} = \C \cup \left\{ \eos \right\}$.
\begin{figure}
     \centering
     \includegraphics[width=0.27\textwidth]{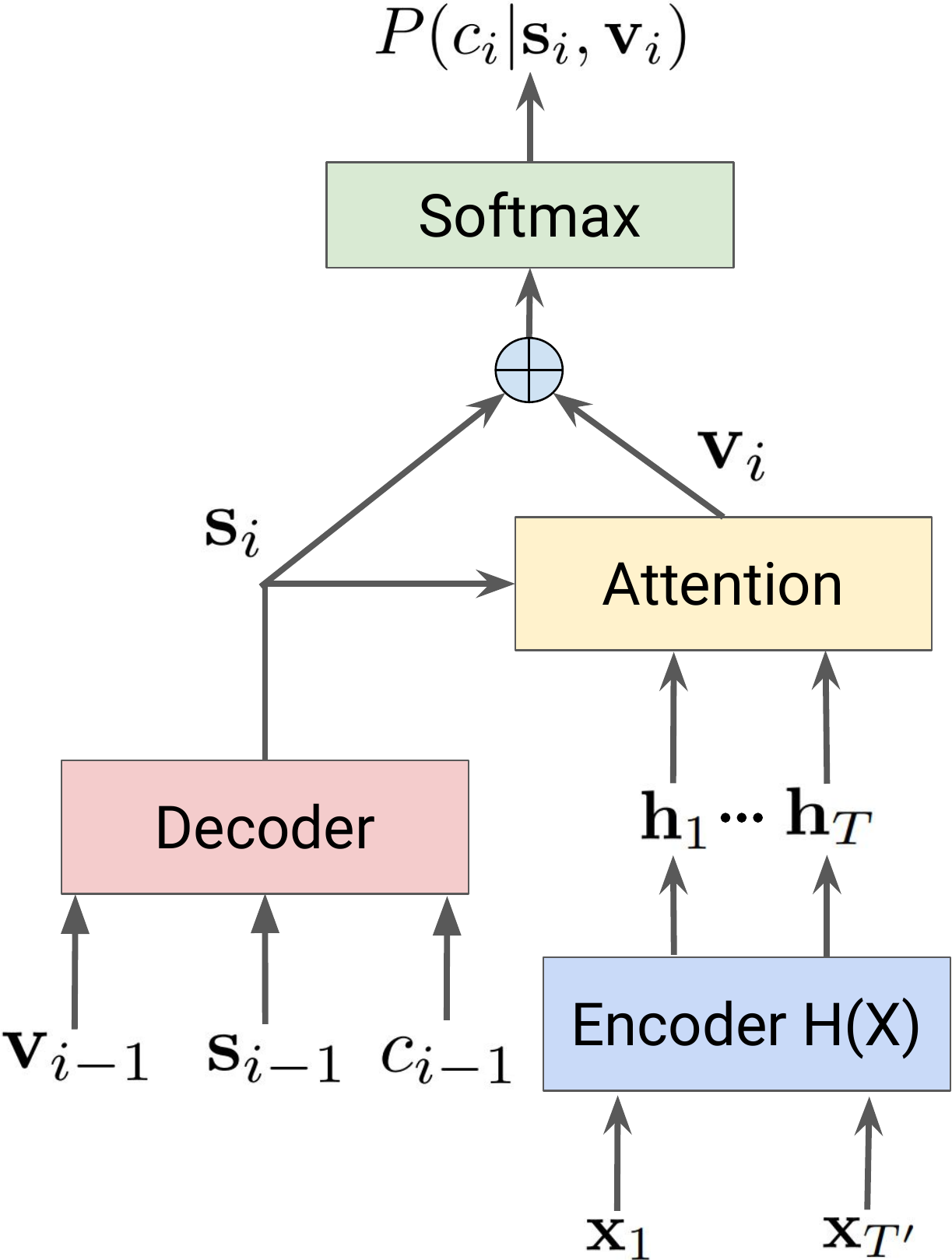}
     \caption{An attention-based encoder decoder (AED) model~\cite{chorowski2015attention, chan2016listen, bahdanau2016end}. The output distribution is conditioned on the decoder state, $\s_i$ (which summarizes the previously decoded symbols), and the context vector, $\cv_i$ (which summarizes the encoder output based on the decoder state). In the seminal work of Chan et al., ~\cite{chan2016listen}, for example, this is accomplished by concatenating the two vectors, as denoted by the $\bigoplus$ symbol in the figure.}
         \label{fig:aed}
\end{figure}
Thus, the AED model, depicted in Figure~\ref{fig:aed}, consists of an encoder network -- which encodes the input acoustic frame sequence, $X = (\x_1, \ldots, \x_{T'})$, into a higher-level representation $H(X) = (\h_1, \ldots, \h_T)$ -- and an attention-based decoder which defines the probability distribution over the set of output symbols, $\C_{\text{eos}}$.
Thus, given a paired training example, $(X, C)$, we denote by $C_e = (c_1, \ldots, c_L, \eos)$,
the ground-truth symbol sequence of length $(L+1)$ augmented with the $\eos$ symbol.
AED models compute the conditional probability of the output sequence augmented with the $\eos$ symbol as:
\begin{align}
P(C_e | X) &= P(C_e | H(X)) \nonumber \\
    &= \prod_{i=1}^{L+1} P(c_i | c_{i-1}, \ldots, c_0 = \sos, H(X)) \nonumber \\
    &= \prod_{i=1}^{L+1} P(c_i | c_{i-1}, \ldots, c_0 = \sos, \cv_i) \nonumber \\
    &= \prod_{i=1}^{L+1} P(c_i | \s_i, \cv_i)\label{eq:las-independence-assumption}
\end{align} 
\noindent where, $\cv_i$ corresponds to a \emph{context vector}, which summarizes the relevant portions of the encoder output, $H(X)$, given the sequence of previous predictions $c_{i-1}, \ldots, c_0$; and, $\s_i$ corresponds to the corresponding decoder state after outputting the sequence of previous symbols, which is produced by updating the decoder state based on the previous context vector and output label:
\begin{equation}
\s_i = \text{Decoder}(\cv_{i-1}, \s_{i-1}, c_{i-1}) \nonumber
\end{equation}
\noindent The symbol $c_0=\sos$ is a special start-of-sentence symbol which serves as the first input to the attention-based decoder before it has produced any outputs.
As can be seen in Eq.~\eqref{eq:las-independence-assumption}, an important benefit of AED models over models such as CTC or RNN-T is that they do not make any independence assumptions between model outputs and the input acoustics, and are thus more general than the implicit alignment models, while being considerably easier to train and implement since we do not have to explicitly marginalize over all possible alignment sequences. 
However, this comes at a cost: previously generated context vectors (which are analogous to the decoded partial alignment in explicit alignment models) are not revised as the decoding proceeds.
Stated another way, while the encoder processing $H(X)$ might be bi-directional, the decoding process in AED models reveals a left-right asymmetry~\cite{mimura18_interspeech}.

\subsubsection{Computing the Context Vector in AED models}
As we mentioned before, the context vector, $\cv_i$, is computed by employing the attention mechanism~\cite{BahdanauChoBengio14}. 
The central idea behind these approaches is to define a state vector $\s_i$ which corresponds to the state of the model after outputting $c_1, \ldots, c_{i-1}$.
The attention function, $\text{atten}(\h_t, \s_i) \in \mathbb{R}$, then defines a score between the model state after outputting $i-1$ previous symbols, and each of the encoded frames in $H(X)$.
These scores can then be normalized using the softmax function to define a set of weights corresponding to each $\h_t$ as:
\begin{equation}
    \alpha_{t, i} = \frac{\exp{\{\text{atten}(\h_t, \s_i)}\}}{\sum_{t'=1}^{T}\exp{\{\text{atten}(\h_{t'}, \s_i)}\}} \nonumber
\end{equation}
\begin{figure}
    \centering
    \includegraphics[width=0.45\textwidth]{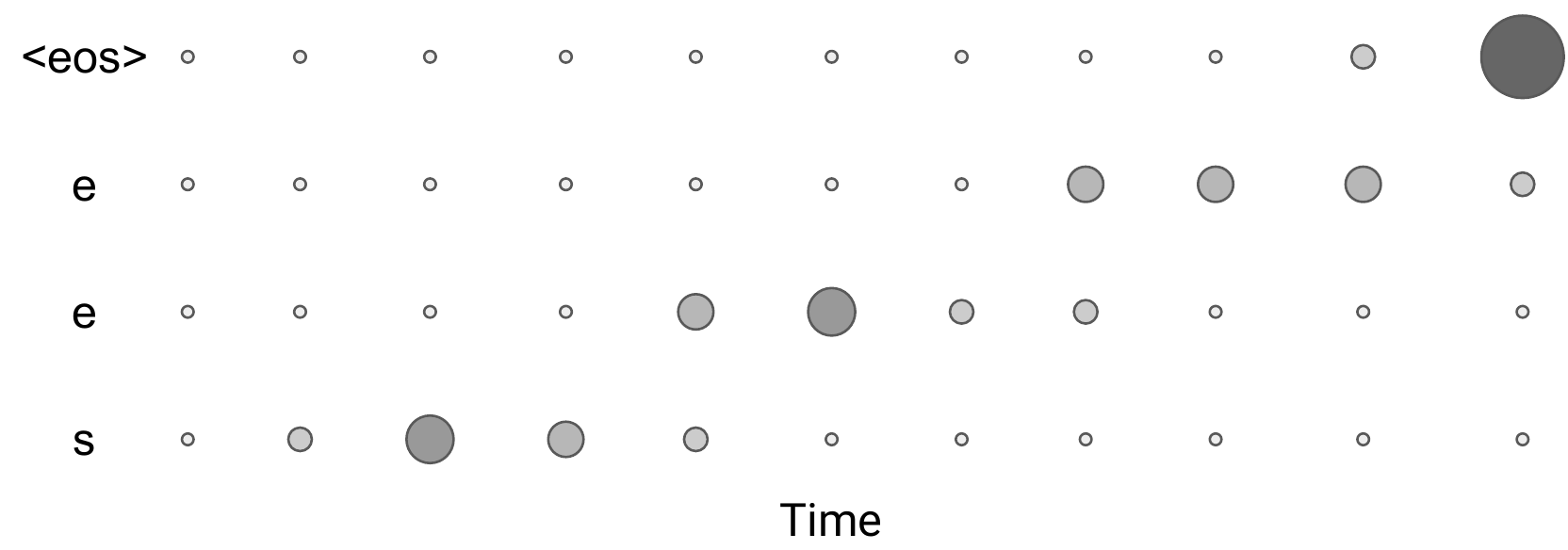}
    \caption{Unlike models such as RNN-T or CTC, AED models do not have explicit alignment. However, it is possible to interpret the attention weights $\alpha_{t, i}$ for a particular output symbol $c_i$ as an alignment weight which is represented above for the target sequence $C = (\texttt{s}, \texttt{e}, \texttt{e}, \texttt{$\eos$})$. In this representation, the size of the circle and the darkness level are proportional to the corresponding attention weights; thus the total probability mass is the same for each row. As illustrated above, the first few frames correspond to the first symbol $c_1 = \texttt{s}$, while the latter frames correspond to the second `e': $c_3 = \texttt{e}$.}
    \label{fig:aed-alignment}
\end{figure}
Intuitively, the weight $\alpha_{t,i}$ represents the relevance of a particular encoded frame $\h_t$ when outputting the next symbol $c_i$, after the model has already output the symbols $c_1, \ldots, c_{i-1}$, as illustrated in Figure~\ref{fig:aed-alignment}.
The context vector summarizes the encoder output based on the computed attention weights:
\begin{equation}
    \cv_{i} = \sum_t \alpha_{t, i} \h_t \nonumber
\end{equation}

A number of possible attention mechanisms have been explored in the literature: the most common forms include dot-product attention~\cite{chan2016listen} and additive attention~\cite{BahdanauChoBengio14}, though a number of other variants have also been explored in the literature (e.g., Gaussian attention~\cite{graves2013generating}, etc.)
Apart from the specific choice of the attention mechanism, a common technique to improve performance involves the use of multiple independent attention heads -- $\cv^{1}_i, \ldots, \cv^{K}_i$ -- which are then concatenated together to obtain the final context vector $\cv_i=\left[\cv^{1}_i; \ldots; \cv^{K}_i\right]$, in the so-called multi-head attention approach~\cite{vaswani2017attention}, or indeed by stacking together multiple attention-based layers in the transformer decoder presented by Vaswani et al.~\cite{vaswani2017attention}.

\subsection{Attention-based E2E Approaches with Alignment Modeling}
AED models, which make no conditional independence assumptions, are extremely powerful, often outperforming explicit attention E2E approaches such as CTC, or RNN-T~\cite{prabhavalkar2017comparison}. 
However, these models also have some significant disadvantages, most notably that the models are fundamentally non-streaming -- i.e., the models must process all acoustic frames before they can generate any output hypotheses.
A somewhat related issue is that the models are extremely sensitive to the length of the acoustic sequences~\cite{chiu2019comparison}, which requires special processing for long-form audio~\cite{chiu2019comparison}.

There have been a number of works that have examined techniques to build streaming AED models.
Generally speaking, these models are motivated by the observation that speech (unlike tasks such as machine translation) exhibits a `local' relationship between the encoded frames (assuming that the encoder is uni-directional) and the output units; thus, unlike the general AED model which computes the context vector, $\cv_i$, as a sum over all input frames $\h_t$, the various proposed models constrain this sum to be computed over a subset of frames to allow for streaming decoding.
In the context of our presentation, it is easiest to think of these models as consisting of an underlying alignment (whether known or unknown) which can be used to perform streaming inference. 

The Neural Transducer (NT)~\cite{JaitlyLeVinyal+s16} explicitly partitions the input encoder frames into $T^W$ non-overlapping chunks of length $W$: $H^W_1 = [\h_1, \ldots, \h_W]; \cdots; H^W_{T^W} = [\h_{T^W + 1}, \ldots, \h_{T^W W}]$, where $T^W =\left\lceil \frac{T}{W} \right\rceil$, and $\h_t = \mathbf{0}$ if $t > T$.
Unlike the AED model which examines all encoded frames when computing the context vector, the NT model is restricted to process a single chunk at a time; the model only advances to the next chunk when it outputs a special end-of-chunk symbol (analogous to $\eos$ in the AED model); inference in the model terminates when the model has output the end-of-chunk symbol in the final chunk $H^W_{T^W}$.
If the alignments of the ground-truth output sequence, $C$, with respect to the $W$-length chunks are unknown, then it is possible to train the system by using a rough initial alignment where symbols are distributed equally among the $T^W$ chunks, followed by iterative refinement by computing the most likely output alignments given the current model parameters~\cite{JaitlyLeVinyal+s16} similar to forced-alignments in HMM-based systems.
An alternate approach~\cite{sainath2018improving} consists of using a separate system (e.g., a classical hybrid system) to get initial alignments (e.g., word-level alignments), which can be used to assign sub-word units to the individual chunks.

An alternative approach, proposed by Raffel et al.~\cite{raffel2017online}, modifies the vanilla AED model by explicitly introducing an alignment module which scans the encoder frames, $H(X)$, from left-to-right to identify whether the current frame should be used to emit any outputs, which is modeled as a Bernoulli random variable. 
If a frame, $\tau$, is selected, then the model produces an output based on the local encoder frame, $\h_\tau$. 
Once the next output is generated, the process is repeated starting from the currently selected frame, thus allowing multiple outputs to be generated at the same frame.
This results in a model which corresponds to hard monotonic alignments between the input speech and the output labels since the models are constrained to generate outputs in a streaming fashion.
A Monotonic Chunkwise Attention (MoChA) model~\cite{chiu2018monotonic} improves upon the work of Raffel et al., by allowing the model to generate the next output using a context vector computed using attention over a local window of frames to the left of the selected frame $\tau$: $\h_{\tau-W+1}, \ldots, \h_{\tau}$.
Thus, the MoChA model consists of a two-level process -- identifying frames where output should be produced following~\cite{raffel2017online}, followed by an AED model over frames to the left of the selected frame.
A refinement to the MoChA model, proposed by Arivazhagan et al.~\cite{arivazhagan2019monotonic} -- the monotonic infinite lookback (MILK) attention model -- computes the context vector over all frames to the left of the selected frame $\tau$ (i.e., $\h_1, \ldots, \h_{\tau}$) at each step.
Another two-fold approach to enable streaming operation is presented in \cite{Moritz+Hori+:2019} under the term of triggered attention, where a CTC-network is used to trigger, i.e.\ control the activation of an AED model with a limited decoder delay.
In \cite{merboldt2019interspeech}, a number of local monotonic attention variants is compared. \cite{zeyer2021:latent-attention} discusses segmental attention variants, while in
\cite{Zeyer+Schmitt+:2023} the related decoding and  the relevance of segment length modeling is investigated, leading to improved generalization towards long sequences. Segmental attention models are related to transducer models \cite{Tian+Yi+:2020}. However, segmental \EtoE\ ASR models are not limited to be realized based on the attention mechanism and may not only be related to a direct HMM \cite{Beck+Hannemann+:2018}, but also have been shown to be equivalent to neural transducer modeling \cite{zhou:segmental-transducer:interspeech2021}.
             
\section{Architecture Improvements to basic E2E models}
\label{sec:improvements}

In this section, we expand on the vanilla E2E models described thus far. Specifically, we  describe various algorithmic changes to these models which are critical for improvement compared to classical ASR systems. First, we describe various ways of combining different complementary E2E models to improve performance. Next, we introduce ways to incorporate context into these models to improve performance on rare proper noun entities. Next, we describe improved encoder and decoder architectures that take better advantage of the many cores on a TPU. Finally, we discuss how to improve the latency of the model through an integrated E2E endpointer. More details about these topics are discussed below.

\subsection{Combinations of Models}
Complementarity exists between different end-to-end models, and there have been numerous attempts that look at combining these methods.

For example, \cite{kim2017joint} found that attention-based models do very poorly in long or noisy utterances, mainly because the model has too much flexibility in predicting alignments since it sees the entire utterance. In contrast, models such as CTC, which have left-to-right constraints during decoding, are much better in these cases. The paper looks at a multi-task learning strategy with both CTC and attention-based models, which gives between a 5-14\% relative improvement in word error rate over attention-based models on a WSJ and Chime task. 

In addition, \cite{SainathPang19} explores combining the benefits of RNN-T and AED. Specifically, RNN-T decodes utterances in a left-to-right fashion, which has also shown improvements for long and noisy utterances. On the other hand, since AED sees the entire utterance, if often shows improvements for utterances where surrounding context is needed to predict the current word, for example, numerics ($1.15$). To combine RNN-T and AED, the authors look at producing a first-pass decode with RNN-T, that is then rescored with AED. To reduce computation, the authors look to share the encoder between RNN-T and AED. The authors find RNN-T+AED gives a 17-22\% relative improvement in word error rate over RNN-T alone on a Voice search task. Other flavors of streaming 1st-pass following by attention-based 2nd-pass rescoring, such as deliberation \cite{hu2020deliberation}, have also been explored.

One of the issues with rescoring is that we are limited to the lattice produced by the 1st-pass system. To address this, methods which run a 2nd-pass beam search have also been explored. To be competitive with a classical-ASR system, the 2nd-pass must be streaming so that it can complete in real time. With this in mind, techniques such as Cascaded Encoder \cite{Arun21}, Y-architecture \cite{Anshuman20} and Universal ASR \cite{Jiahui20}, are such examples. 

\subsection{Incorporating Context}

Contextual biasing to a specific domain, including a user’s song names, app names and contact names, is an important component of any production-level automatic speech recognition (ASR) system. Contextual biasing is particularly challenging in E2E models because these models keep a small list of candidates during beam search, and also do poorly on proper nouns, which is the main source of biasing phrases. There have been a few approaches in the literature to incorporate context.

One approach is to construct a stand-alone weighted finite state transducer (FST) representing the biasing phrases. The scores from the biasing FST are interpolated with the scores of the \EtoE\ model during beam search, with special care taken to ensure we do not over- or under-bias phrases. This technique is known as shallow-fusion contextual biasing \cite{ding2019biasing}.

Another approach looks at injecting biasing phrases into the model in an all-neural way. For example, \cite{pundak2018deepcontext} represents a set of biasing phrases by embedding vectors. These vectors are fed as additional input to an attention-based model, which can then choose to attend to the phrases and hence boost the chances of predicting the phrases. Furthermore, \cite{Kim2018DialogContextAE} looked at biasing towards dialog context in an all-neural way. In addition, ~\cite{bruguier2019phoebe} looks at using phoneme and grapheme information to embed biasing phrases.
Finally, \cite{delcroix2018auxiliary} uses an utterance-wise context vector like an i-vector computed by a pooling across frame-by-frame hidden state vectors obtained from a sub network (this sub network is called a sequence-summary network).

\subsection{Encoder/Decoder Structure}
\label{sec:enc_dec}
There have been improvements to encoder architectures of \EtoE\ models over time. The first end-to-end models used long short-term memory recurrent neural networks (LSTMs), for both the encoder and decoder. The main drawback of these sequential models is that each frame depends on the computation from the previous frame, and therefore multiple frames cannot be batched in parallel. 

With the improvement of hardware, specifically on-device Edge Tensor Processing Units (TPUs), with thousands of cores, architectures that can better take advantage of the hardware, have been explored. Such architectures include convolution-based architectures, such as ContextNet \cite{Han+Zhang+:2020}. The use of self-attention to replace the sequential recurrence in LSTMs was explored in Transformers for ASR \cite{dong2018speech,Zhang2020}. Finally, combining self-attention with convolution, known as Conformer \cite{Gulati+Qin+:2020conformer}, or multi-layer perceptron \cite{peng2022branchformer}, was also explored. Both Transformer and Conformer have shown competitive performance to LSTMs on many tasks \cite{karita2019comparative,guo2021recent}. 

On the decoder side, research for transducer models has shown that a large LSTM decoder can be replaced with a simple embedding lookup table, that attends to only a few previous tokens from the model \cite{VarianiRybachAllauzen+20,Rami21,ghodsi2020rnn,zhou:phoneme-transducer:2021, prabhavalkar2021less}. This demonstrates that most of the power of the E2E model is in the encoder, which has been a consistent theme of both \EtoE\ as well as classical hybrid HMM models.

\subsection{Integrated Endpointing}

An important characteristic of streaming speech recognition systems is that they must endpoint quickly, so that the ASR result can be finalized and sent to the server for the appropriate action to be performed. Within this, endpointing is typically done with an external voice-activity detector. Since endpointing is both an acoustic and language model decision, recent works in streaming RNN-T models \cite{chang19unified, Li2020} have looked to predict a microphone closing token $\eos$ at the end of the utterance. For example, our sentence may be ``What's the weather $\eos$". These models have shown improved latency and WER tradeoff by having the endpointing decision predicted as part of the model. Furthermore,  \cite{yoshimura2020end,fujita21b_interspeech} explored using the CTC blank symbol for endpoint detection.
    
\section{Training \EtoE\ Models}
\label{sec:training}

In general, training of \EtoE\ models follows deep learning schemes
\cite{Bengio:2012,Schmidhuber:2015}, with specific consideration of
the sequential structure and the latent alignment problem to be
handled in ASR. \EtoE\ ASR models may be trained end-to-end,
notwithstanding potential elaborate training schedules and extensive
data augmentation. One of the appeals of end-to-end models is that
they do not assume conditional independence between the input
frames. Given a training set $\mathcal{T} = \{(X_i, C_i)\}_{i=1}^N$,
the training criterion $\Ls$ to be minimized (which is equivalent to
maximizing the total conditional log-likelihood) can be written as:
$\Ls = \sum_{i=1}^{N} \log{P(C_i | X_i)}$.

\subsection{Alignment in Training}

\EtoE\ models such as RNN-T and CTC introduce an additional blank
token $\blank$ for alignment. Therefore optimization implies
marginalizing across all alignments, which requires the
forward-backward algorithm~\cite{Baum:1972,Rabiner+Juang:1986} for
efficient computation of the training criterion and its gradient, with
minor modifications for CTC, RNN-T, and RNA models, as well as
classical (full-sum) hybrid ANN/HMMs corresponding to the differences
in alignments defined in each of these models. In comparison, AED
models do not have a latent variable for explicit alignment. We refer
the interested reader to the individual papers for further details on
the training
algorithms~\cite{bengio1991neural,haffner1993connectionist,
  graves2006connectionist, graves2012sequence, graves2013speech,
  sak2017recurrent, chorowski2015attention, chan2016listen,
  bahdanau2016end}.

As part of the training procedure, all \EtoE\ as well as classical
hidden Markov models for ASR provide mechanisms to solve the
underlying sequence alignment problem - either explicitly via
corresponding latent variables, as in CTC, RNN-T or RNA, and also
hybrid ANN/HMM, or implicitly, as in AED models. Also, the distinction
between speech and silence needs to be considered, which may be
handled explicitly by introducing silence as a latent label (hybrid
ANN/HMM), or implicitly by not labeling silence at all, as currently
is the standard in virtually all \EtoE\ models.

\EtoE\ models also may take advantage of hierarchical training
schedules. These schedules may comprise several separated training
passes and explicit, initially generated aligments that are kept fixed
for some Viterbi-style \cite{Bellman:1957, Viterbi:1967, ney1984use}
training epochs before reenabling {\EtoE}-style full-sum training that
marginalizes over all possible alignments. Such an alternative
approach is employed by Zeyer et al.~\cite{zeyer2020new}, where an initial full-sum RNN-T model is used to generate an alignment and continue with framewise cross-entropy training. This greatly simplifies the training process by replacing the
summation over all possible alignments
in Eq.~\eqref{eq:rna-independence-assumption} by a single term
corresponding to the alignment sequence generated. Recently, a
similar procedure has been introduced in \cite{Zhou+Michel+:2022}
also employing \EtoE\ models, only. In this work, CTC is used to initialize
the training and to generate an initial alignment, followed by
intermediate Viterbi-style RNN-T training and final full-sum fine
tuning. This process showed improved convergence compared to
full-sum-only training approaches.

It is interesting to note that in contrast to the RNN-T and RNA
label-topologies, CTC does not require alignments with single label
emissions per label position. However, training CTC models eventually
does lead to single label emissions per hypothesized label. An
analysis of this property of CTC training which is usually called
\textit{peaky behavior} can be found in \cite{zeyer2021:peakyctc} and
references therein. \cite{Laptev+Majumdar+:2022} even introduces a CTC variant without non-blank loop transitions.

\subsection{Training with External Language Models}

\EtoE\ ASR models generally are normalized on sequence
level. Therefore, sequence training with the maximum mutual
information criterion \cite{He:2008} is the same as standard cross
entropy/conditional likelihood training. However, once external
language models are included in the training phase, sequence
normalization needs to be included explicitly, leading to MMI sequence
discriminative training. This has been exploited as a further approach
to combine \EtoE\ models with external language models trained on
text-only data already in the training phase
\cite{zeineldeen2021:ilm,Wynands+Michel+2022:approx-recombination,Yang+Zhou+:2022}.

\subsection{Minimum Error Training}

Since the objective of speech recognition is to minimize word error
rate (WER), there has been a growing number of research studies that
incorporate this into the objective function. These methods, known as
sequence/discriminative learning, have shown great improvements for
classical ASR
\cite{Valtchev1997,Povey2001,schlueter01:cod,Kingsbury2009,heigold2012:DTfundamentals},
and have since been explored in \EtoE\ models.  Typically these losses
are constructed by running in 'beam-search' mode rather than
teacher-forcing mode, and construct a loss from the errors made from
the candidate hypotheses in the beam. Thus, this type of training
first requires training the model to optimize $P(C|X)$ so that we can
learn a good set of weights to run a beam search. However, also direct
approaches have been introduced that avoid this separation to train
discriminatively from scratch
\cite{Povey+Peddinti+:2016,Michel+Schlueter+:2019}.

Papers that explore penalizing word errors include, Minimum Word Error
Rate (MWER) training \cite{prabhavalkar2018mwer}, where the loss
function is constructed such that the expected number of word errors
are minimized. Further work includes MWER for RNN-T and
self-attention-T \cite{weng2019minimum}, as well as MWER using prefix
search instead of n-best \cite{baskar2019promising}. Also, there have
been studies that consider MWER in terms of reinforcement learning
\cite{tjandra2018sequence,karita2018sequence}. Finally, Optimal
Completion Distillation (OCD) \cite{Sabour+Chan+:arXiv2018} looks to
minimize the total edit distance using an efficient dynamic
programming algorithm. Finally, another body of research with sequence
training introduce a separate external language model at training
time, \cite{Michel+Schlueter+2020:lm_integration}, which can also be
done efficiently via approximate lattice recombination
\cite{Wynands+Michel+2022:approx-recombination} and also lattice-free
approaches~\cite{Yang+Zhou+:2022}.

\subsection{Pretraining}

All \EtoE\ models as well as classical hidden Markov models for ASR
provide holistic models that in principle enable training from
scratch. However, many strategies exist to initialize and guide the
training process to reach optimal performance and/or to obtain
efficient convergence by applying pretraining and model growing
\cite{hinton06,bengio2006greedy}. Supervised layer-wise pretraining
has been successfully applied for classical
\cite{Seide+:2011,zeyer17:lstm}, as well as attention-based ASR models
\cite{zeyer2018:returnn}, which can be combined with intermediate
subsampling schemes \cite{zeyer2018improved}, and model growing
\cite{zeyer2018neurips-irasl}. Pretraining approaches utilizing
untranscribed audio, large-scale semi-supervised data and/or
multilingual data
\cite{wave2vec-original,wave2vec-journal,Scanzio+Laface+:2008,tuske2013:multilingual,Zhou+Xu+:2018,Adams+Wiesner+:2019,Hou+Dong+:2020,Pratap+Sriram+:2020,Li+Pang+:2021,Zhang+Park+:2022BigSSL,Chen+Zhang+:2022MAESTRO,Radford+Kim+:2022Whisper}
would deserve a self-contained survey and they are applicable for
hybrid DNN/HMM and \EtoE\ approaches likewise - they will not be
further discussed here.

\subsection{Training Schedules and Curricula}

Dedicated training schedules have been developed to guide the
optimization process and as part of that reach proper alignment
behavior explicitly or implicitly
\cite{zeyer2018improved,zeyer2020new,Zhou+Michel+:2022}.

Many approaches exist for learning rate control
\cite{Vogl+Mangis+:1988accelerating,Keskar+Saon:2015nonmonotone}:
NewBob \cite{Renals+Morgan+:1991,Johnson+Ellis+:2004quicknet} and
enhancements \cite{Keskar+Saon:2015nonmonotone}, global vs.\
parameter-wise learning rate control (exponential decay, power decay,
etc.) \cite{Senior+Heigold+:2013empirical}, learning rate warmup
\cite{vaswani2017attention}, warm restarts/cosine
annealing \cite{Loshchilov+Hutter:2017}, weight decay vs. gradually
decreasing batch size \cite{Smith+Kindermans+:2017}, fine-tuning
\cite{Howard+Ruder:2018} or population-based training
\cite{Jaderberg+Dalibard:2017}. For a survey of meta learning also
cf.\ \cite{Hospedales+Antoniou+:2020}.

Sequence learning approaches also consider curriculum learning
\cite{Elman:1993,Bengio+Louradour+:2009}, e.g.\ by considering short
sequences first \cite{amodei2016deep,Tuske+Saon+:2020}, interim
increase of subsampling \cite{zeyer2018improved} initially more
subsampling, or for multi-speaker ASR training sort mixed speech by
SNR and start with speakers of balanced energy and mixed gender
\cite{zhang2020improving}.

\subsection{Optimization and Regularization}

Optimization usually is based on stochastic gradient descent
\cite{Polyak:1964}, with momentum
\cite{Nesterov:1983,Sutskever+Martens+:2013importance}, and a number
of corresponding adaptive approaches, most prominently Adam
\cite{Kingma+Ba:2015Adam} and variants thereof
\cite{Kingma+Ba:2015Adam,zeyer17:lstm,Tuske+Saon+:2021}.

Investing more training epochs seems to provide improvements
\cite[Table 8]{zeyer2020new}, and also averaging over epochs has been
reported to help \cite{karita2019comparative}. For a discussion of the
double descent effect and its relation to the amount of training data,
label noise and early stopping cf.\ \cite{Nakkiran+Kaplun+:2019}.

Regularization strongly contributes to training performance, incl.\
L2 and weight decay \cite{Krogh+Hertz:1991,Loshchilov+Hutter:2017},
weight noise \cite{Murray+Edwards:1994}, adaptive mean L1/L2 \cite{Graves:2011},
gradient noise \cite{Neelakantan+Vilnis+:2015},
dropout \cite{Hinton+Srivastava+:2012,Krizhevsky+Sutskever+:2012,Gal+Ghahramani:2016},
layer dropout \cite{Huang+Sun+:2016,pham2019very,lee2021intermediate},
dropconnect \cite{Wan+Zeiler+:2013},
zoneout \cite{Krueger+Maharaj+:2016},
smoothing of attention scores \cite{chorowski2015attention}, 
label smoothing \cite{Szegedy+Vanhoucke+:2016},
scheduled sampling \cite{Bengio+Vinyals+:2015},
auxiliary loss \cite{Szegedy+Vanhoucke+:2016, Trinh+Dai+:2018auxiliary},
variable backpropagation through time \cite{Williams+Peng:1990,Merity+Keskar+:2018}, mixup \cite{meng2021mixspeech}, increased frame rate \cite{Tuske+Saon+:2021},
or batch normalization \cite{Ioffe+Szegedy:2015}.

\subsection{Data Augmentation}
\label{sec:data_aug}

Training of \EtoE\ ASR models also benefit from data augmentation
methods, which might also be viewed as regularization
methods. However, their diversity and impact on performance justifies
a separate overview.

Most data augmentation methods perform data perturbation by exploiting
certain dimensions of speech signal variation: speed perturbation
\cite{Kanda+Takeda+:2013,Ko+Peddinti+:2015}, vocal tract length
perturbation \cite{Jaitly+Hinton:2013,Kanda+Takeda+:2013}, frequency
axis distortion \cite{Kanda+Takeda+:2013}, sequence noise injection
\cite{Saon+Tuske+:2019}, SpecAugment
\cite{park2019specaugment}, or semantic mask
\cite{Wang+:2020semantic}. Also, text-only data may be used to
generate data using text-to-speech (TTS) on feature
\cite{hayashi2018back} or signal level
\cite{Rossenbach+Zeineldeen+:2021}. In a comparison of the effect of
TTS-based data augmentation on different \EtoE\ ASR architectures in
\cite{Rossenbach+Zeineldeen+:2021}, AED seemed to be the only
architecture that took advantage of data generated by TTS,
significantly.

In a recent study \cite{Tuske+Saon+:2020} and corresponding follow-up
work \cite{Tuske+Saon+:2021}, many of the regularization and data
augmentation methods listed here have been exploited jointly leading
to state-of-the-art performance on the Switchboard task for a
single-headed AED model.
            
\section{Decoding \EtoE\ Models}
\label{sec:decoding}

This section mainly describes several decoding algorithms for end-to-end speech recognition.
The basic decoding algorithm of end-to-end ASR tries to estimate the most likely sequence $\hat{C}$ among all possible sequences, as follows:
\begin{align}
    \hat{C} 
    & = \arg \max _{C \in \mathcal{U}^{\ast}} P(C | X) \nonumber
\end{align}
where $^{\ast}$ means a Kleene closure to represent a set of all possible sequences composed of token vocabulary $\mathcal{U}$.
$P(C | X)$ is a probabilistic distribution function obtained by an E2E ASR system.
The following section describes how to obtain the recognition result $\hat{C}$.

\subsection{Greedy search}
The Greedy search algorithm is mainly used in CTC, which ignores the dependency of the output labels as follows:
\begin{align}
    \hat{A} 
    & = \prod_{t=1} ^T \left( \arg \max _{a_t} P(a_t|X) \right) \nonumber 
\end{align}
where $a_t$ is an alignment token introduced in Section \ref{sec:ctc}.
The original character sequence is obtained by converting alignment token sequence $\hat{A}$ to the corresponding token sequence $\hat{C}$.
The argmax operation can be performed in parallel over input frame $t$, yielding fast decoding \cite{graves2006connectionist,graves2012connectionist}, although the lack of the output dependency causes relatively poor performance than the attention and RNN-T based methods in general.

CTC's fast decoding is further boosted with transformer \cite{vaswani2017attention,dong2018speech,karita2019comparative} and its variants \cite{Gulati+Qin+:2020conformer,guo2021recent} since their entire computation across the frame is parallelized \cite{pham2019very,higuchi2020mask}. 
For example, the non-autoregressive models, including Imputer \cite{chan2020imputer}, Mask-CTC \cite{higuchi2020mask}, Insertion-based modeling \cite{fujita20_interspeech}, Continuous integrate-and-fire (CIF) \cite{dong2020cif} and other variants \cite{nozaki21_interspeech,higuchi2021comparative} have been actively studied as an alternative non-autoregressive model to CTC.
\cite{higuchi2021comparative} shows that CTC greedy search and its variants achieve ~0.06 real-time factor (RTF: (decoding time) / (input length)) by using Intel(R) Xeon(R) Silver
4114 CPU, 2.20GHz.
The paper also shows that the degradation of the non-autoregressive models from the attention/RNN-T methods with beam search is not significantly large (19.7\% with self-conditioned CTC \cite{nozaki21_interspeech} vs. 18.5/18.9\% with attention/RNN-T).

The greedy search algorithm is also used as approximate decoding for the attention, RNA, CTC, and RNN-T based methods as follows:
\begin{align}
    \hat{c}_{i}
    & = \arg \max _{c_i} P(c_i|\hat{C}_{1:i-1}, X) \text{ for } i = 1, \dots, N \nonumber \\
    \hat{a}_{t}
    & = \arg \max _{a_t} P(a_t|\hat{A}_{1:t-1}, X) \text{ for } t = 1, \dots, T \nonumber 
\end{align}
The greedy search algorithm does not consider alternate hypotheses in a sequence compared with the beam search algorithm described below.
However, it is known that the degradation of the greedy search algorithm is not so large \cite{chan2016listen,sak2017recurrent}, especially when the model is well trained in matched conditions\footnote{On the other hand, in the AED models, increasing the search space does not consistently improve the speech recognition performance \cite{chorowski2017towards,Zhou+Schlueter+:Interspeech2020}, which is also observed in neural machine translation \cite{koehn2017six}.}.

\subsection{Beam search}
The beam search algorithm is introduced to approximately consider a subset of possible hypotheses $\tilde{\mathcal{C}}$ among all possible hypotheses $\mathcal{U}^{\ast}$ during decoding, i.e., $\tilde{\mathcal{C}} \subset \mathcal{U}^{\ast}$ .
A predicted output sequence $\hat{C}$ is selected among a hypothesis subset $\tilde{\mathcal{C}}$ instead of all possible hypotheses $\mathcal{U}^{\ast}$, i.e.,
\begin{equation}
    \hat{C} = \arg \max _{C \in \tilde{\mathcal{C}}} P(C|X)
    \label{eq:beam_search_start}
\end{equation}
The beam search algorithm is to find a set of possible hypotheses $\tilde{\mathcal{C}}$, which can include promising hypotheses efficiently by avoiding the combinatorial explosion encountered with all possible hypotheses $\mathcal{U}^{\ast}$.

There are two major beam search categories: 1) \textit{frame synchronous} beam search and 2) \textit{label synchronous} beam search.
The major difference between them is whether it performs hypothesis pruning for every input frame $t$ or every output token $i$.
The following sections describe these two algorithms in more detail.

\subsection{Label synchronous beam search}
\label{sec:label_sync_beam}
Suppose we have a set of partial hypotheses up to $(i-1)$th token $\tilde{\mathcal{C}} _{1:i-1}$.
A set of all possible partial hypotheses up to $i$th token $\mathcal{C} _{1:i}$ is expanded from $\tilde{\mathcal{C}} _{1:i-1}$ as follows:
\begin{equation}
    \mathcal{C} _{1:i} = \{(\tilde{\mathcal{C}} _{1:i-1}, c_i = c) \}_{c \in \mathcal{U}}
    \label{eq:label_partial_hyp}
\end{equation}
The number of hypotheses $|\mathcal{C} _{1:i}|$ would be $|\tilde{\mathcal{C}} _{1:i-1}| \times |\mathcal{U}|$ at most.
The beam search algorithm prunes the low probability score hypotheses from $\mathcal{C} _{1:i}$ and only keeps a certain number (beam size $\Delta$) of hypotheses at $i$ among $\mathcal{C} _{1:i}$.
This pruning step is represented as follows:
\begin{equation}
    \tilde{\mathcal{C}} _{1:i} = \mathrm{NBEST}_{C_{1:i} \in \mathcal{C} _{1:i}} P(C_{1:i}|X), \text{ where } |\tilde{\mathcal{C}} _{1:i}| = \Delta
    \label{eq:label_beam}
\end{equation}
Note that $\mathrm{NBEST}(\cdot)$ is an operation to extract top $\Delta$ hypotheses in terms of the probability score $P(C_{1:i}|X)$ computed from an end-to-end neural network, or a fusion of multiple scores described in Section \ref{sec:fusion}.

In the label synchronous beam search, the length of the output sequence ($N$) is not given.
Therefore, during this pruning process, we also add the hypothesis that reaches the end of an utterance (i.e., predict the end of sentence symbol $\eos$) to a set of hypotheses $\tilde{\mathcal{C}}$ in Eq.~\eqref{eq:beam_search_start} as a promising hypothesis.  

Since the label synchronous beam search does not explicitly depend on the alignment information, sequence hypotheses of the same length might cover a completely different number of encoder frames, unlike the frame synchronous beam search, as pointed out by \cite{zhou:segmental-transducer:interspeech2021}.
As a result, we observe that the scores of very short and long segment hypotheses often become the same range, and the beam search wrongly selects such hypotheses.
\cite{hori2017joint} shows an example of such extreme cases, which results in large deletion and insertion errors for short and long-segment hypotheses, respectively.
Thus, the label synchronous beam search requires heuristics to limit the output sequence length to avoid extremely long/short output sequences.
Usually, the minimum and maximum length thresholds are determined proportionally to the input frame length $|X|$ with tunable parameters $\rho _{\text{min}}$ and $\rho _{\text{max}}$ as $L _{\text{min}}  = \lfloor \rho _{\text{min}} |X| \rfloor, L _{\text{max}} = \lfloor \rho _{\text{max}} |X| \rfloor$.
Although these are quite intuitive ways to control the length of a hypothesis, the minimum and maximum output lengths depend on the token unit or type of script in each language.
The other heuristics is to provide the additional score related to the output length or attention weights, including the length penalty and the coverage term \cite{tu2016modeling,chorowski2017towards}.
The end-point detection \cite{hori2018end} is also used to estimate the hypothesis length automatically.
\cite{Zhou+Schlueter+:Interspeech2020} redefines the implicit length model of the attention decoder to take into account beam search, resulting in consistent behavior without degradation for increasing beam sizes.

The label synchronous beam search algorithms of CTC are realized by marginalizing all possible alignments for each label hypothesis \cite{graves2006connectionist}.

\subsection{Frame synchronous beam search}
In contrast to the label synchronous case in Eq.~\eqref{eq:label_beam}, the frame synchronous beam search algorithm performs pruning at every input frame $t$, as follows:
\begin{equation}
    \tilde{\mathcal{C}} _{1:i(t)} = \mathrm{NBEST}_{C_{1;i(t)}} P(C_{1;i(t)}|X), \text{ where } |\tilde{\mathcal{C}} _{1:i(t)}| = \Delta \nonumber
\end{equation}
where $C_{1;i(t)}$ is an $i(t)$-length label sequence obtained from the alignment $A _{1:t}$, which is introduced in Section~\ref{sec:explicit-alignment-e2e-approaches}.
$P(C_{1;i(t)}|X)$ is obtained by summing up all possible alignments $A _{1:t} \in \mathcal{A}_{(X, C_{1;i(t)})}$. 
A set of alignments $\mathcal{A}$ depends on explicit alignment approaches, including CTC, RNN-T, and RNA.
$\mathcal{C} _{1:i(t)}$ is an expanded partial hypotheses up to input frame $t$, similar to Eq.~\eqref{eq:label_partial_hyp}.

Compared with the label synchronous algorithm, the frame synchronous algorithm needs to handle additional output token transitions inside the beam search algorithm.
The frame synchronous algorithm can be easily extended in online and/or streaming decoding thanks to the explicit alignment information with input frame and output token.

Classical approaches to beam search for HMM, but also CTC and RNN-T variants, are based on weighted finite state transducers (WFST) \cite{mohri2002weighted,hori2013speech,miao2015eesen} or lexical prefix trees \cite{Haeb-Umbach+Ney:1994,Ney+Ortmanns:2000,zhou:phoneme-transducer:2021}. They are categorized as frame synchronous beam search.
These methods are often combined with an N-gram language model or a full-context neural language model \cite{Hori+Kubo+:2014,Beck+Zhou+:2019}.
RNN-T \cite{graves2012sequence,saon2020alignment} and CTC prefix search \cite{hannun2014first} can deal with a neural language model by incorporating the language model score in the label transition state.
Triggered attention approaches \cite{moritz2019triggered,moritz2020streaming} allow us to use AED models in frame-synchronous beam search together with CTC and neural LM, which applies on-the-fly rescoring to the hypotheses given by CTC prefix search using the AED and LM scores.

\subsection{Block-wise decoding}
Another beam search implementation uses a fixed-length block unit for the input feature.
In this block processing, we can use the future context inside the block by using the non-causal encoder network based on the BLSTM, output-delayed unidirectional LSTM, or transformer (and its variants).
This future context information avoids the degradation of the fully causal network.
In this setup, the chunk size becomes the trade-off of controlling the latency and accuracy.
This technique is used in both
RNN-T \cite{jain2019rnn,yeh2019transformer,lu20g_interspeech} and AED \cite{chiu2018monotonic,wang21ba_interspeech,tsunoo2021streaming}.
Block-wise processing is especially important for AED since it can provide block-wise monotonic alignment constraint between the input feature and output label, and realize block-wise streaming decoding.

\subsection{Model fusion during decoding}
\label{sec:model_fusion_during_decoding}
Similar to the classical HMM-based beam search, we combine various scores obtained from different modules, including the main end-to-end ASR and LM scores.

\subsubsection{Synchronous score fusion}
The most simple score fusion is performed when the scores of multiple modules are synchronized.
In this case, we can simply add the multiple scores at each frame $t$ or label $i$.
The most well-known score combination is the LM shallow fusion.
\subsubsection*{LM shallow fusion}
As discussed in Section \ref{sec:lm}, various neural LMs can be integrated with end-to-end ASR.
The most simple integration is based on LM shallow fusion \cite{hwang2017character}\cite{hori2017advances}\cite{kannan2018analysis}, as discussed in Section \ref{sec:shallow_fusion}, which (log-)linearly add the LM $P _{\mathsf{lm}}(C_{1:i})$ to e2e ASR scores $P(C_{1:i}|X)$ during beam search in Eq.~\eqref{eq:label_beam} as follows:
\begin{align}
\log P(C_{1:i}|X) \rightarrow \log P(C_{1:i}|X) + \gamma \log P _{\mathsf{lm}} (C_{1:i}) \nonumber 
\end{align}
where $\gamma$ is a language model weight.
Of course, we can combine the other scores, such as the length penalty and coverage terms, as discussed in Section \ref{sec:label_sync_beam}.
\subsubsection{Asynchronous score fusion}
If we combine the frame-dependent score functions $P(a_t|\cdot)$ (e.g., CTC, RNN-T) and label-dependent score functions $P(c_i|\cdot)$ (e.g., attention, language model).
We have to deal with the mismatch between the frame and label time indexes $i$ and $t$.

In the time-synchronous beam search, this fusion is performed by incorporating the language model score in the label transition state \cite{CollobertPuhrschSynnaeve16,he2019streaming,saon2021advancing}.
\cite{hannun2014first} also combines a word-based language model and token-based CTC model by incorporating the language model score triggered by the word delimiter (space) symbol.

In the label-synchronous beam search, we first compute the label-dependent scores from the frame-dependent score function by marginalizing all possible alignments given a hypothesis label sequence.
CTC/attention joint decoding \cite{hori2017joint} is a typical example, where the CTC score was computed by marginalizing all possible alignments based on the CTC forward algorithm \cite{graves2012connectionist}.
This approach eliminates the wrong alignment issues and difficulties of finding the correct end of sentences in the label-synchronous beam search \cite{hori2017joint,watanabe2017hybrid}.

Note that the model fusion method during beam search can realize simple one-pass decoding, while it limits the time unit of the models to be the same or it requires additional dynamic programming to adjust the different time units, especially for the label-synchronous beam search.
This dynamic programming computation becomes significantly large when the length of the utterance becomes larger and requires some heuristics to reduce the computational cost \cite{seki2019vectorized}.

\subsection{Lexical constraint during score fusion}
Classically, we use a word-based language model to capture the contextual information with the word unit, and also consider the word-based lexical constraint for ASR.
However, end-to-end ASR often uses a letter or token unit and it causes further unit mismatch during beam search.
As described in previous sections, the classical approach of incorporating the lexical constraint from the token unit to the word unit is based on a WFST.
This method first makes a TLG transducer composed of the token (T), word lexicon (L), and word-based language transducers (G) \cite{miao2015eesen}.
This TLG transducer is used for both CTC \cite{miao2015eesen} and attention-based \cite{bahdanau2016end}.

Another approach used in the time synchronous beam search is to  insert the word-based language model score triggered by the word delimiter (space) symbol \cite{hannun2014deep}.
To synchronize the word-based language model with a character-based end-to-end ASR, \cite{hori2017multi} combines the word and character-based LMs with the prefix tree representation, while \cite{hori2018end,wang2019espresso} uses look-ahead word probabilities to predict next characters instead of using the character-based LM.
The prefix tree representation is also used for the sub-word token unit case \cite{tuske2019advancing,drexler2019subword}. 

\subsection{Multi-pass fusion}

The previous fusion methods are performed during the beam search, which enables the one-pass algorithm.
The popular alternative methods are based on multi-pass algorithms where we do not care about the synchronization and perform n-best or lattice scoring by considering the entire context within an utterance.
\cite{chan2016listen} uses the N-best rescoring techniques to integrate a word-based language model.
\cite{mimura18_interspeech} combines forward and backward searches within a multi-pass decoding framework to combine bidirectional LSTM decoder networks.
Recently two-pass algorithms of switching different end-to-end ASR systems have been investigated, including RNN-T $\rightarrow$ attention \cite{sainath2019two} CTC $\rightarrow$ attention \cite{zhang2021wenet,wu2021u2++}.
This aims to provide the streamed output in the first pass and the re-scoring with attention in the second pass refines the previous output to satisfy the real-time interface requirement and high recognition performance.

In addition to the N-best output in the above discussion, there is a strong demand of generating a lattice output for better multi-pass decoding thanks to richer hypothesis information in a lattice.
The lattice output can also be used for spoken term detection, spoken language understanding, and word posteriors.
However, due to the lack of Markov assumptions, RNN-T and AED cannot merge the hypothesis and cannot generate a lattice, unlike the HMM-based or CTC systems.
To tackle this issue, there are several studies of modifying these models by limiting the output dependencies in the fixed length (i.e.,  finite-history) \cite{zapotoczny2019lattice,VarianiRybachAllauzen+20}, or keeping the original RNN-T structure but merging the similar hypotheses during beam search \cite{prabhavalkar2021less}.

\subsection{Vectorization across both hypotheses and utterances}
We can accelerate the decoding process by vectorizing multiple hypotheses during the beam search, where we replace the score accumulation steps for each hypothesis with vector-matrix operations for the vectorized hypotheses. 
It has been studied in RNN-T \cite{he2019streaming,saon2021advancing,kim2020accelerating} and attention-based \cite{seki2019vectorized}.
This modification allows us to take advantage of the parallel computing capabilities of multi-core CPUs and GPUs, resulting in significant speedups and also enabling us to process multiple utterances in a batch simultaneously.

Major deep neural network and end-to-end ASR toolkits support this vectorization.
For example, Tensorflow\footnote{https://www.tensorflow.org/api\_docs/python/tf/contrib/seq2seq/BeamSearchDecoder}\cite{abadi2016tensorflow}, and  FAIRESEQ\footnote{https://github.com/pytorch/fairseq/blob/master/fairseq/sequence\_generator.py}\cite{ott2019fairseq} provide a vectorized beam search interface for a generic sequence to sequence task, and it can be used for attention-based end-to-end ASR.
The end-to-end ASR toolkit including ESPnet\footnote{https://github.com/espnet/espnet}\cite{seki2019vectorized}, ESPRESSO\footnote{https://github.com/freewym/espresso}\cite{wang2019espresso}, and RETURNN\footnote{https://github.com/rwth-i6/returnn}\cite{doetsch2017returnn} also support the vectorized beam search algorithm.
       
\section{LM Integration}
\label{sec:lm}

Language models (LMs) have long been used for ASR in combination with acoustic models to improve the recognition accuracy \cite{jelinek1997statistical}.
The main role of the LM is to help the ASR system generate the most probable sentence hypothesis in terms of language.
Practically, an LM is designed as a probabilistic model such as an $N$-gram model \cite{jelinek1997statistical} or a neural network LM \cite{Bengio+:2000,mikolov2010recurrent}, which can be trained with a large number of example sentences.
The LM provides a probability distribution $P(C)$ over a random variable of sentence, $C$, which is used to score sentence hypotheses and encourage the system to choose a more likely hypothesis. 

As mentioned in Section \ref{sec:relationship}, hybrid ASR systems typically utilize a separate LM in the inference stage, while end-to-end (\EtoE) systems basically utilize a single ASR model including a network component that plays a role of LM, where the entire network has been trained jointly. For example, the prediction network of RNN-T and the decoder network of AED models take on the role of LM. Therefore, \EtoE\ ASR does not seem to require separate LMs anymore. Nevertheless, many studies have demonstrated that separate LMs help improve the recognition accuracy in \EtoE\ ASR.

There are presumably three reasons that \EtoE\ ASR still requires an external LM:
\paragraph{Compensation for poor generalization}
\EtoE\ models need to learn a more complicated mapping function than classical modular-based models such as acoustic models. Consequently, \EtoE models tend to face overfitting problems if the amount of training data is not sufficient. Pretrained LMs potentially compensate for the less generalized predictions made by \EtoE\ models. 
\paragraph{Use of external text data}
\EtoE\ models need to be trained using paired speech and text data, while LMs can be trained with only text data. Generally, text data can be collected more easily than the paired data. The training speed of an LM is also faster than that of \EtoE\ models for the same number of sentences. Accordingly, the LM can be improved more effectively with external text data, providing additional performance gain to the ASR system.
\paragraph{Domain adaptation}
Domain adaptation helps improve recognition accuracy when the \EtoE\ model is applied to a specific domain. However, domain adaptation of the \EtoE\ model requires a certain amount of paired data in the target domain. Also, when multiple domains are assumed, it may be costly to maintain multiple \EtoE\ models for the domains the system supports. If a pre-trained LM for the target domain is available, it may more easily improve recognition accuracy for domain-specific words and speaking styles without updating the \EtoE\ model.

This section reviews various types of LMs used for \EtoE\ ASR and fusion techniques to incorporate them into the main \EtoE\ network.

\subsection{Language models}
An LM provides a prior probability distribution $P(C)$.
If sentence $C$ can be decomposed into a sequence of tokens such as characters, subwords, and single words, the probability distribution can be computed based on the chain rule as:
\begin{equation}
    P(C)=\prod_{i=1}^{L+1} P(c_i|c_{0:i-1}) \nonumber 
\end{equation}
where $c_i$ denotes the $i$-th token of $C$, and $c_{0:i-1}$ represents token sequence $c_0, c_1, \dots, c_{i-1}$, assuming $c_0=\sos$ and $c_{L+1}=\eos$. 

Most LMs are designed to provide the conditional probability $P(c_i|c_{0:i-1})$, i.e., they are modeled to predict the next token given a sequence of the preceding tokens.
We briefly review such LMs here.

\subsubsection{N-gram LM}
$N$-gram LMs have long been used for ASR  \cite{jelinek1997statistical}.
Early \EtoE\ systems in \cite{miao2015eesen,bahdanau2016end,chorowski2017towards} also employed an $N$-gram LM.
The $N$-gram models rely on the Markov assumption that the probability distribution of the next token depends only on the previous $N-1$ tokens, i.e., $P(c_i|c_{0:i-1}) \approx P(c_i|c_{i-N+1:i-1})$, where $N$ is typically 3 to 5 for word-based models and a larger number for subword- and character-based models.
The maximum likelihood estimates of $N$-gram probabilities are determined based on the counts of $N$ sequential tokens in the training data set as
\begin{equation}
    P(c_i|c_{i-N+1:i-1})=\frac{\mathcal{K} (c_{i-N+1},\dots,c_i)}{\sum_{c_i}\mathcal{K} (c_{i-N+1},\dots,c_i)} \nonumber 
\end{equation}
where $\mathcal{K}(\cdot)$ denotes the count of each token sequence.
Since the data size is finite, it is important to apply a smoothing technique to avoid estimating the probabilities based on zero or very small counts for rare token sequences. Those techniques compensate the $N$-gram probabilities with lower order models, e.g., $(N-1)$-gram models, according to the magnitude of the count \cite{chen1996empirical}.

The advantage of the $N$-gram models is their simplicity, although they underperform state-of-the-art neural LMs. In the training, the main step is to just count the $N$ tuples in the data set, which is required only once. In the decoding, the LM probabilities can be obtained very quickly by table lookup or can be attached to a decoding graph, e.g., WFST, in advance.

\subsubsection{FNN-LM}
The feed-forward neural network (FNN) LM was proposed in \cite{Bengio+:2000}, which estimates $N$-gram probabilities using a neural network. The network accepts $N-1$ tokens, and predicts the next token as
\begin{align}
    P(c_i|&c_{i-N+1:i-1})=\mathrm{softmax}(W_o h_i + b_o) \nonumber \\
        h_i & = \mathrm{tanh}(W_h e_i + b_h) \nonumber \\
        e_i & = \mathrm{concat}(E(c_{i-N+1}), \dots, E(c_{i-1})) \nonumber 
\end{align}
where $W_o$ and $W_h$ are weight matrices, and $b_o$ and $b_h$ are bias vectors.
$E(y)$ provides an embedding vector of $c$, and $\mathrm{concat}(\cdot)$ operation concatenates given vectors \footnote{We omit the optional direct connection from the embedding layer to the softmax layer in \cite{Bengio+:2000} for simplicity}.
This model first maps each input token to an embedding space, and then obtains hidden vector $h_i$ as a context vector representing the previous $N-1$ tokens.
Finally, it outputs the probability distribution of the next token through the $\mathrm{softmax}$ layer. Although this LM still relies on the Markov assumption, it outperforms classical $N$-gram LMs described in the previous section.
The superior performance of FNN-LM is primarily due to the distributed representation of each token and context. The LM learns to represent token/context vectors such that semantically similar tokens/contexts are placed close to each other in the embedding and context spaces. 
Since this representation has a better smoothing effect than the count-based one used for $N$-gram LMs, FNN-LM can provide a better generalization than $N$-gram LMs for predicting the next token.
Neural network-based LMs basically utilize this type of representation. 

\subsubsection{RNN-LM}
A recurrent neural network (RNN) LM was introduced to exploit longer contextual information over $N-1$ previous tokens using recurrent connections \cite{mikolov2010recurrent}. Unlike FNN-LM, the hidden vector is computed as
\begin{align}
    h_i & = \mathrm{recurrence}(e_i, h_{i-1}) \nonumber \\
    e_i & = E(c_{i-1})\nonumber 
\end{align}
where $\mathrm{recurrence}(e_i, h_{i-1})$ represents a recursive function, which accepts previous hidden vector $h_{i-1}$ with input $e_i$, and outputs next hidden vector $h_i$.
In the case of simple (Elman-type) RNN, the function can be computed as \begin{align}
    \mathrm{recurrence}(e, h)=\mathrm{tanh}(W_h e + W_r h + b_h) \nonumber 
\end{align}
where $W_r$ is a weight matrix for the recurrent connection, which is applied to the previous hidden vector $h$.
This recurrent loop makes it possible to hold the contextual information in the hidden vector without limiting the context to $N-1$ tokens.
However, the past information decays exponentially as tokens are processed with this recursion. Therefore, currently stacked LSTM layers are more widely used for the recurrence part, which have separate internal memory cells and gating mechanisms to keep the long-range contextual information \cite{sundermeyer2012lstm}.
With this mechanism, RNN-LMs outperform other $N$-gram-based models in many tasks.

\subsubsection{ConvLM}
ConvLM \cite{dauphin2017language} replaces the recurrent connections used in RNN-LMs with gated temporal convolutions.
The hidden vector is computed as
\begin{align}
    h_i =& h_i' \otimes \sigma(g_i) \nonumber \\
    h_i' = & e_{i-k+1:i} * W + b \nonumber \\
    g_i = & e_{i-k+1:i} * V + c \nonumber 
\end{align}
where $\otimes$ is element-wise multiplication $*$ is a temporal convolution operation, and $k$ is the patch size. $\sigma(g_i)$ represents a gating function of convoluted activation $h_i'$, where $\sigma(\cdot)$ is a sigmoid function.
$W$ and $V$ are matrices for convolution and $b$ and $c$ are bias vectors. The convolution and gating blocks are typically stacked multiple times with residual connections. In  \cite{zeghidour2018fully}, a ConvLM with 14 blocks has been applied for \EtoE\ ASR. 
The advantage of ConvLM is its computational efficiency, achieving similar performance to that of RNN-LMs, even with a fixed context size consisting of short tokens such as letters and word pieces. Specifically, its highly parallelizable architecture is useful to train the model with a large training data set.

\subsubsection{Transformer LM}
Transformer architecture \cite{vaswani2017attention} has been applied to LMs \cite{al2019character} and used for ASR \cite{irie2019interspeech,karita2019comparative}, where the LMs are designed as a Transformer decoder without any inputs from other modules such as encoders. The hidden vector is computed as
\begin{align}
    h_i =& \mathrm{FFN}(h_i') + h_i' \nonumber \\
    h_i'=& \mathrm{MHA}(e_i, e_{1:i}, e_{1:i}) + e_i \nonumber
\end{align}
where $\mathrm{FFN}(\cdot)$ and $\mathrm{MHA}(\cdot,\cdot,\cdot)$ denote a feed forward network and a multi-head attention module, respectively. The multihead attention and feed-forward blocks are typically stacked multiple times, e.g., 6 times \cite{karita2019comparative}, to obtain the final hidden vector. 
The advantage of Transformer LMs is that they can take all previous tokens into account through the self-attention mechanism without summarizing them into a fixed-size memory like RNN-LMs.
Thus, the long previous context can be fully utilized to predict the next token, achieving better performance than RNN-LMs.
However, the computational complexity increases quadratically as the length of the sequence. Therefore, the length of the context is typically limited to a fixed size or within every single sentence.
To overcome this limitation, Transformer-XL \cite{dai2019transformer} reuses already computed activations, which includes information on farther previous tokens, and the model is trained with a truncated back-propagation through time (BPTT) algorithm. Compressive Transformer \cite{rae2020compressive} extends this approach to utilize even longer contextual information by incorporating a compression step to keep older but important information in a fixed-size memory network.

\subsection{Fusion approaches}
\label{sec:fusion}
There are several ways to incorporate LMs into \EtoE\ ASR. Researchers have investigated various LM fusion approaches to improve ASR accuracy in different situations.

\subsubsection{Shallow fusion}
\label{sec:shallow_fusion}
Shallow fusion is the most popular approach to combine the pre-trained \EtoE\ model and LM in the inference time. As we described in Section \ref{sec:model_fusion_during_decoding},
 the shallow fusion simply combines the \EtoE\ and LM scores by a log-linear combination as
\begin{equation}
Score(C|X) = \log P(C|X) + \gamma \log P(C)
\label{eq:lm_shallow_fusion}
\end{equation}
where $\gamma$ is a scaling factor for the LM    \cite{hwang2017character}\cite{hori2017advances}\cite{kannan2018analysis}.
The advantage of this approach is that it is easy and effective when there are no major differences between the source and target domains.

\subsubsection{Deep fusion}
Deep fusion \cite{gulcehre2015using} is an approach to combine an LM with an \EtoE\ model using a joint network.
All the network parameters are trained jointly so that the models collaborate better to improve the recognition accuracy, where the joint network is used to combine the \EtoE\ and LM states through a gating mechanism that controls the contribution of the LM according to the current state.
 
\subsubsection{Cold fusion}
Cold fusion \cite{sriram2018cold} is another approach to combine the LM like Deep fusion, but the \EtoE\ model is learned with a pretrained LM while freezing the LM parameters. 
Since the \EtoE\ model is aware of the LM throughout training, it learns to use the LM to reduce language specific information and capture only the relevant information to map the source to the target sequence.
This mechanism reduces the role of LM in the \EtoE\ model and alleviates the language bias of the training data.
Accordingly, the \EtoE\ model becomes more robust to domain mismatches between the training data and the target domain.
Unlike Deep fusion, Cold fusion makes it possible to combine the \EtoE\ model with a pretrained LM for the target domain, improving the recognition accuracy.
Component fusion \cite{shan2019component} extends Cold fusion to use a pretrained LM with transcriptions of the training data for the \EtoE\ model, more focusing on reducing the bias of the training data.

\subsubsection{Internal LM estimation}
There is another approach to reduce language bias in training data through Shallow fusion.
The language bias is a problem when a big domain mismatch exists between the source domain (training data) and the target domain (test data) because the \EtoE\ model scores are strongly dependent on the language priors in the source domain.
To remove such a bias from the score, we can explicitly estimate the LM that represents the language priors, called {\it Internal LM}, and subtract the LM score from the ASR score of Eq.~\eqref{eq:lm_shallow_fusion}, i.e.,
\begin{equation}
Score(C|X) = \log P_\varphi(C|X) - \gamma_\varphi \log P_\varphi(C) + \gamma_\tau \log P_\tau(C) \nonumber
\end{equation}
where subscripts $\varphi$ and $\tau$ indicate the source and target domains, respectively. $\gamma_\varphi$ and $\gamma_\tau$ are their scaling factors.
Subtracting the internal LM score corresponds to approximating acoustic probability density $P_\varphi(X|C)$ because $P_\varphi(X|C) \propto P_\varphi(C|X)/P_\varphi(C)$ is satisfied for fixed $X$, where the ASR score can be seen as a classical hybrid ASR system. Accordingly, the subtracted \EtoE\ model score plays a role of acoustic model and makes it more domain independent in terms of language, achieving a higher recognition accuracy in combination with the external LM $P_\tau(C)$.

The density ratio method \cite{McDermott+Sak+:2020} trains an internal LM using the transcript of the training data. Hybrid autoregressive transducer (HAT) \cite{VarianiRybachAllauzen+20} extends RNN-T so that the model becomes the internal LM when the encoder output is eliminated, i.e., set to zero. This approach simplifies the framework by utilizing the prediction network as the internal LM, which avoids training an additional LM and using it in the inference time. In the work of \cite{Meng+Parthasarathy+:2020}, an approach similar to HAT has been proposed where the internal LM is formulated on top of standard RNN-T and attention-based encoder-decoder models, respectively. 
In \cite{zeineldeen2021:ilm}, several techniques to estimate internal LMs have been proposed for AED models, where an estimated bias vector is fed to the LM instead of a zero vector. The bias vector can be estimated by averaging encoder states or context vectors, or by a small LSTM predicting the context vector based on the decoder label context, only. These techniques to estimate the internal LM were also evaluated for RNN-T in \cite{Zhou+Zheng+:2022}.

\subsection{Use of large-scale pretrained LMs}
In recent years, LMs trained with large-scale text data are available for different NLP tasks. BERT \cite{devlin2019bert} and GPT-2 \cite{radford2019language} are representative models based on Transformer LMs. Such LMs have also been applied to \EtoE\ ASR systems in different ways, e.g., N-best rescoring \cite{salazar2020masked} and dialog context embedding \cite{kim2019gated}.
             
\section{Relationship To classical ASR}
\label{sec:relationship}

\EtoE\ systems provide a number of properties that set them apart from
classical ASR systems.  Ideally, \EtoE\ systems define ASR models
that integrate all knowledge sources into a single global joint model
that does not utilize secondary knowledge sources and avoids the
classical separation into acoustic and language models. These global
joint models are completely trained from scratch using a single global
training criterion based on a single kind of (transcribed) training
data and thus require less ASR domain-specific knowledge, provided 
sufficient amounts of training data are available. Similarly, 
recognition may be performed in a single pass and without the use of 
an explicit lexicon.

\EtoE\ systems share some of their properties with classical ASR
systems, while other \EtoE\ properties are sacrificed for reasons of
performance and depending on the amount of training data available.
In the following, we will discuss various relationships that can be
observed between \EtoE\ and classical ASR approaches.

First of all, \EtoE\ models introduce alternative alignment modeling
approaches to ASR. While the attention mechanism provides a
qualitatively novel approach to map acoustic observation sequences to
label sequences, transducer approaches
\cite{graves2006connectionist,graves2012sequence,sak2017recurrent,CollobertPuhrschSynnaeve16}
handle the alignment problem in a way that can be interpreted to be
similar to HMMs with a specific model topology, including
marginalization over
alignments~\cite{haffner1993connectionist,Zeyer+Beck+:Interspeech2017,Raissi+Zhou+:2023}. CTC
models can also be employed in an HMM-like fashion during
decoding~\cite{miao2015eesen}. Moreover, transducer approaches are equivalent to segmental models/direct HMM~\cite{zhou:segmental-transducer:interspeech2021}.

The most prominent property shared between \EtoE\ and classical
statistical ASR systems is the single-pass decoding strategy,
integrating all knowledge sources involved (models, components) before
performing a final decision~\cite{ney1984use}. This includes the use
of full label context dependency both for \EtoE\ systems
\cite{graves2012connectionist,tripathi2019monotonic,chorowski2017towards,Hannun+Lee+:2019,Tuske+Saon+:2020,tuske2019advancing,Li+Liu+:2019,Bahar+Makarov+:2020},
as well as classical systems via full-context language
models~\cite{Huang+Zweig+:2014,Hori+Kubo+:2014,Beck+Zhou+:2019,Jorge+Gimenez+:2019}.
In classical ASR systems even HMM alignment path summation may be
retained in search~\cite{Zhou+Schlueter+:2020}.  Both \EtoE\ as well
as classical ASR systems employ beam search in decoding. However,
compared to classical search approaches, \EtoE\ beam search usually is
highly simplified with very small beam sizes around $1$ to $100$
\cite{chorowski2015attention,chan2016listen,chorowski2017towards,zeyer2018improved}.
Very small beam sizes also partly mask a length bias exhibited by
\EtoE\ attention-based encoder-decoder models
\cite{Sountsov+Sarawagi:2016,Murray+Chiang:2018}, thus trading model
errors against search errors \cite{Stahlberg+Byrne:2019}. An overview
of approaches to handle the length bias beyond using small beam sizes
in ASR is presented in \cite{Zhou+Schlueter+:Interspeech2020}.

Another prominent feature of \EtoE\ systems is their direct
character-level modeling avoiding phoneme-based modeling and
pronunication lexica
\cite{GravesJaitly14,hannun2014deep,miao2015eesen,chan2016listen,Lu+Zhang+:ICASSP2016,chorowski2017towards,Zhang+Chan+:ICASSP2017,Toshniwal+Tang+:arXiv2017,Renduchintala+Ding+:arXiv2018,Sabour+Chan+:arXiv2018,Weng+Cui+:Interspeech2018},
with some even heading for whole-word modeling
\cite{Lu+Zhang+:ICASSP2016,soltau2016neural}. However,
character-level modeling also is viable with classical hybrid HMM
architectures \cite{Le+Zhang+:ASRU2019_Chenones} and has been shown to
work even with standard HMM models w/o neural networks
\cite{Kanthak+Ney:2002}.

Many classical ASR search paradigms are based on multipass
approaches that successively generate search space representations
applying increasingly complex acoustic and/or language models
\cite{Deshmukh+Ganapathiraju+:1999,Nguyen+Schwartz:1999,Ney+Ortmanns:2000}.
However, multipass strategies also are employed using \EtoE\ models,
which however softens the \EtoE\ concept. 
Decoder model combination is pursued in a two-pass approach,
while even retaining latency constraints as in
\cite{SainathPang19}. Further multipass approaches incl.\ \EtoE\
adaptation approaches
\cite{Sari+Moritz+:2020,Weninger+Andres-Ferrer+:2019,Meng+Gaur+:2019,Tomashenko+Esteve:2019}.

The architecture of classical ASR systems provides a separation
into acoustic model and language model. In contrast to this,
\EtoE\ models avoid this separation and define a joint model. While
this allows for training with a single objective, it limits training
of the (implicit) prior to the transcriptions of the audio training
data. To exploit further text-only training data, usually a separate
LM is combined with \EtoE\ models, nonetheless. However, due to the
implicit prior of \EtoE\ models, i.e.\ the internal language model, combination with separate language
models is not straightforward and requires corresponding internal language model estimation and compensation
approaches, e.g.\
\cite{McDermott+Sak+:2020,VarianiRybachAllauzen+20,Meng+Parthasarathy+:2020,zeineldeen2021:ilm,zeyer2021:transducer-librispeech}.
At least from the recognition accuracy perspective, it remains
unclear, if the clear separation of acoustic modeling and language
modeling in the classical ASR architecture is a disadvantage
because of separate training objectives, or rather an advantage, since
text-only training data may be used easily. Also, the language model
training objective, i.e.\ language model perplexity, is observed to
correlate well with word error rate
\cite{Bahl+Jelinek+:1983,Makhoul+Schwartz+:PNAS1995,Klakow+Peters:SpeechCommunication2002,Sundermeyer+Ney+:ASLP2015}.
Furthermore, discriminative approaches to language modeling
\cite{Tachioka+Watanabe:2015,Hori+Hori+:2016} may be viewed as a step
towards joint modeling.

While standard hybrid ANN/HMM training for ASR using frame-wise cross
entropy already is discriminative, it is not yet sequence
discriminative, requires prior alignments and also lacks consideration
of an (external) language model during training. However, these potential 
shortcomings may be remedied by using sequence discriminative training 
criteria \cite{He:2008} and lattice-free training approaches \cite{Povey+Peddinti+:2016}.

In contrast to strict \EtoE\ systems, the classical ASR
architecture includes the use of secondary knowledge sources beyond
the primary training data, i.e. (transcribed) speech audio for
acoustic model training, and textual data for language model
training. Most prominently, this includes the use of a pronunciation
lexicon and the definition of a phoneme set. Secondary resources like
pronunciation lexica may be helpful in low-resource
scenarios. However, their generation often is costly and may even
introduce errors, like pronunciations from a lexicon not reflecting
the actual pronunciations observed. Therefore, for large enough
training resources, secondary knowledge sources might become obsolete
\cite{sainath2018lexicon}, or even harmful, in case of erroneous
information introduced \cite{Wooters+Stolcke:1994,McGraw+Badr+:2013}.

Classical ASR models usually are trained successively, with
knowledge derived from models trained earlier injected into later
training stages, e.g.\ in the form of HMM state alignments. However,
such approaches from classical ASR might also be interpreted as
specific training schedules. Initializing deep learning models using
HMM alignments obtained from acoustic models based on mixtures of
Gaussians may be interpreted in this way, with the Gaussian mixtures
serving as an initial shallow model. In classical ASR, also
approaches training deep neural networks from scratch while avoiding
intermediate training of Gaussians has been proposed
\cite{Senior+Heigold+:2014,Gosztolya+Grosz+:2016,Hadian+Sameti+:2018},
also in combination with character-level modeling
\cite{Le+Zhang+:ASRU2019_Chenones}. Another step towards more
integrated training of classical systems has been to apply
discriminative training criteria avoiding intermediate (usually
lattice-based) representations of competing word sequences
\cite{Soltau+Kingsbury+:2005,Povey+Peddinti+:2016,Hadian+Povey+:2018,Kanda+Fujita+:2018,Michel+Schlueter+:2019}.

The training of classical ASR systems usually applies secondary
objectives to solve subtasks like phonetic clustering. The
classification and regression trees (CART) approach is used to cluster
triphone HMM states \cite{Breiman,Young+Woodland:1993}. More recent
approaches proposed clustering within a neural network modeling
framework, while still retaining secondary clustering objectives
\cite{Wiesler+Heigold+:2010,Gosztolya+Grosz+:2016}. However, also in
\EtoE\ approaches secondary objectives are used, most prominently for
subword generation, e.g.\ via byte-pair
encoding~\cite{Sennrich+:2015}. Also, available pronunciation lexica
can be utilized indirectly for assisting subword generation for \EtoE\
systems \cite{Xu+Ding+:2019,zhou:ADSM:interspeech2021}, which are
shown to outperform byte-pair encoding. Within classical ASR
systems, phonetic clustering also can be avoided completely by
modeling phonemes in context directly~\cite{raissi21:tchham}.

It is interesting to observe that specifically attention-based
encoder-decoder models require larger numbers of training epochs to
reach high performance, e.g.\ for a comparison of systems trained on
Switchboard 300h cf.\ Table 5 in \cite{zeineldeen20:phon-att}. 
Also, attention-based encoder-decoder models have been shown to suffer
from low training
resources~\cite{luescher2019interspeech,Park+Zhang+:2020}, which can
be improved by a number of approaches, including regularization
techniques \cite{Tuske+Saon+:2020} as well as data augmentation using
SpecAugment \cite{ParkChanZhangChiuEtAl19} and text-to-speech (TTS)
\cite{baskar2019semisupervised}. SpecAugment also is shown to improve
classical hybrid HMM models~\cite{zhou:tedlium2:icassp2020}. TTS on
the other hand considerably improved attention-based encoder-decoder
models trained on limited resources, but did not reach the performance
of other \EtoE\ approaches or hybrid HMM models, which in turn were
not considerably improved by TTS \cite{Rossenbach+Zeineldeen+:2021}.
Multilingual approaches also help improve ASR development for low
resource tasks, again both for classical
\cite{Cui+Kingsbury+:2015}, as well as for \EtoE\ systems
\cite{adams2019massively,Kannan+Datta+:2019}.

Speech recognition research has always been pushed by international
evaluation campaigns (e.g. NIST) and corresponding benchmark tasks.
The competition between classical and \EtoE\ approaches is nicely
reflected in the widely used Librispeech \cite{Librispeech} and
Switchboard \cite{Switchboard} tasks, showing that \EtoE\ models gain
momentum. On Librispeech, current best published classical hybrid
systems range around 2.3\% (test-clean) and 4.9\% (test-other) word
error rate \cite{wang2020transformer,luescher2019interspeech}, while
there already are a number of \EtoE\ systems providing similar
performance
\cite{ParkChanZhangChiuEtAl19,park2019specaugment,synnaeve2019end,Wang+:2020semantic},
with some \EtoE\ systems clearly outperforming former state-of-the-art
results with word error rates down to 1.8\% (test-clean) and 3.7\%
(test-other) \cite{Kim+Wu+:2023} with similar results reported in \cite{Gulati+Qin+:2020conformer,Han+Zhang+:2020}.  Merging insights from classical HMM-based and monotonic RNN-T provided similarly well results with a limited training budget \cite{Zhou+Michel+:2022}. Finally,
when trained on Switchboard 300h, the current best result, obtained with
an \EtoE\ system \cite{Tuske+Saon+:2021} is $5.4\%$ compared to
$6.6\%$ word error rate for the best hybrid system
result~\cite{kitza19:interspeech} on the HUB5'00 Switchboard test set.

\section{Overall performance trends of E2E approaches in common benchmarks}
\label{sec:comparison}

\begin{figure}
     \centering
     \includegraphics[width=1.\columnwidth]{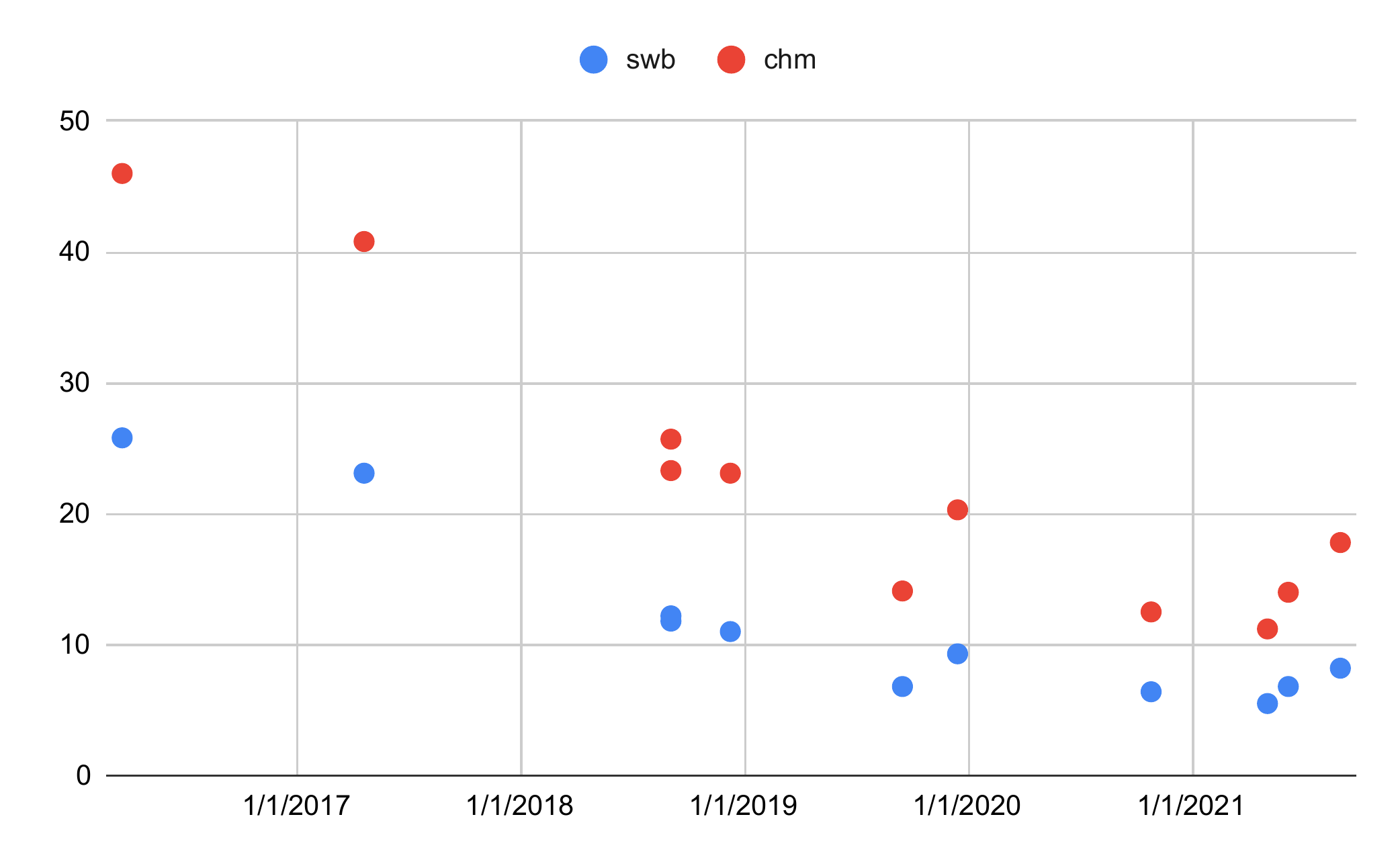}
     \caption{E2E ASR performance improvement in the switchboard task.}
     \label{fig:swbd}
\end{figure}

\begin{figure}
     \centering
     \includegraphics[width=1.\columnwidth]{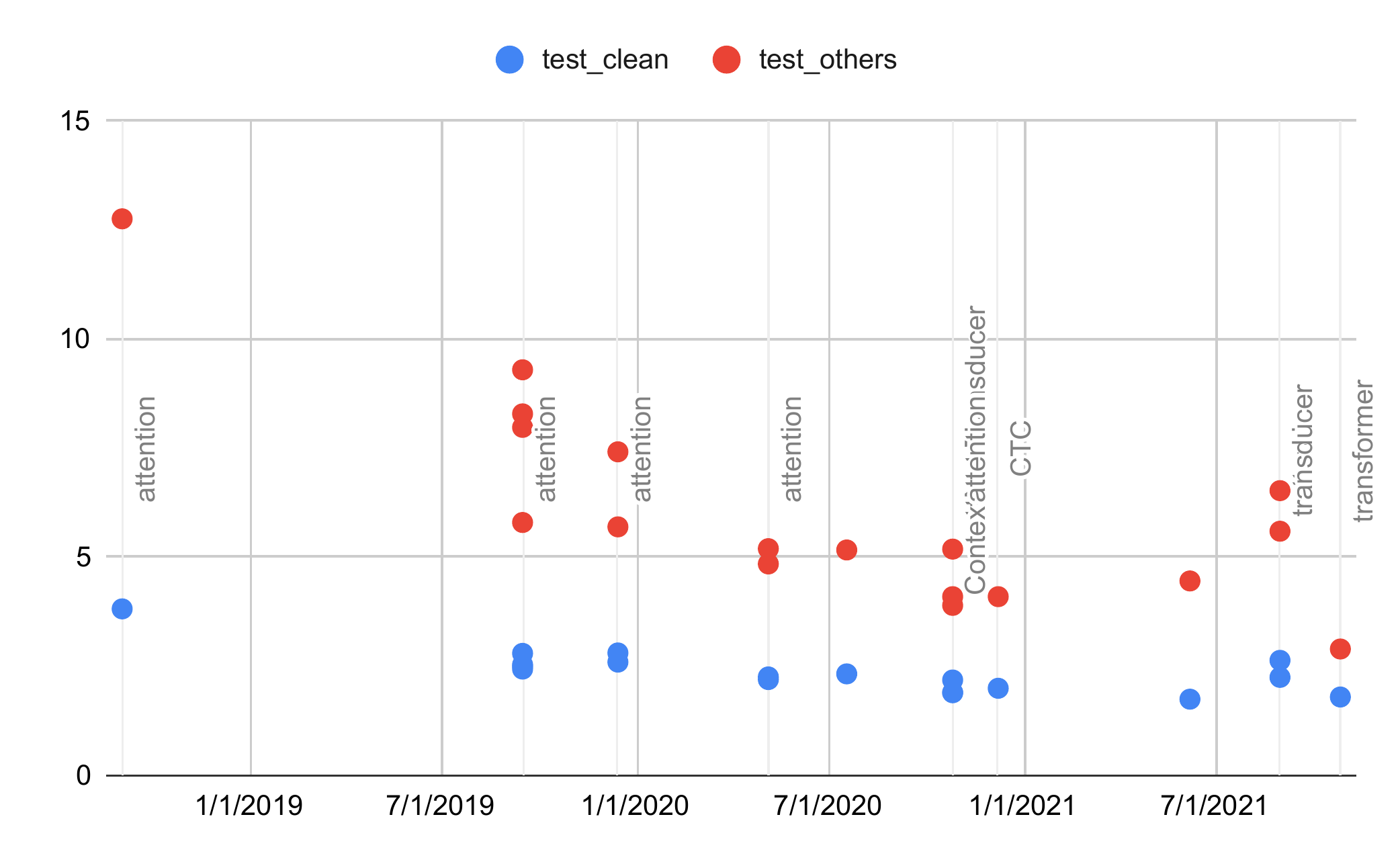}
     \caption{E2E ASR performance improvement in the Librispeech task.}
     \label{fig:libri}
\end{figure}

This section summarizes various techniques with the common ASR benchmarks based on Librispeech \cite{Librispeech} in Figure~\ref{fig:libri} and switchboard (SWBD) \cite{Switchboard} in Figure~\ref{fig:swbd} to see the trajectory of the techniques developed in end-to-end ASR.
We choose these two databases because they are widely used in speech and machine learning communities and cover spontaneous (SWBD) and read speech (Librispeech) speaking styles.
Figures~\ref{fig:swbd} and~\ref{fig:libri} show that the performance improvement from the initial trials \cite{Toshniwal+Tang+:arXiv2017,zeyer2018improved} based on the E2E models is significant, and the error rates of all tasks become less than half\footnote{For readers who want to know the latest update of these benchmarks can also check \url{https://github.com/syhw/wer_are_we} and \url{https://github.com/thu-spmi/ASR-Benchmarks/blob/main/README.md}.}.

Although the overall trends show that the ASR performance was steadily improved as time goes, there are several remarkable gains.
One significant gain observed in both benchmarks in the middle of 2019 comes from the data augmentation method represented by SpecAugment \cite{park2019specaugment,Wang+:2020semantic}, as discussed in Section~\ref{sec:data_aug}.
The subsequent gains mostly come from the exploration of the new E2E architectures, including transformer \cite{karita2019comparative,zeyer2019comparison}, conformer \cite{Gulati+Qin+:2020conformer,guo2021recent}, and contextnet \cite{Han+Zhang+:2020} on top of SpecAugment, as discussed in Section~\ref{sec:enc_dec}.
Such an exploration is also performed in language modeling to improve the ASR performance \cite{irie2019interspeech,karita2019comparative}.
The final gain observed in the Librispeech benchmark in 2021 is based on self-supervised learning \cite{Baevski2020wav2vec2,hsu2021hubert} and semi-supervised learning  \cite{synnaeve2019end,ng2021pushing}.
These techniques utilize a considerable amount of unlabeled in-domain speech data (e.g., Libri-light 60K hours \cite{kahn2020libri}).

\section{Deployment of E2E models}
\label{sec:applications}

Many of the ideas discussed in this paper have been explored by various industry research labs \cite{ChiuSainathWuPrabhavalkarEtAl18, KimSamsung2019, HsiaoCanNgEtAl20, ShiWangWuEtAl21, ChenWuWangEtAl21} inter alia.
In this section, we review the development of on-device production-level systems at Google as a typical case study for the deployment. 
The first streaming E2E model, deployed to production, was launched in 2019 for the Pixel 4 smartphone~\cite{he2019streaming,Sainath2020}. This model used a streaming RNN-T first pass system, while re-scoring first-pass hypotheses with an AED system in the second pass.
In addition, FST-based contextual biasing~\cite{ding2019biasing} was employed in the model, which was critical to obtain accurate results for diverse queries.
This model ran on CPU and was much faster than real time.

In 2020, for the Pixel 5 smartphone~\cite{bo21system}, the system was improved further to reduce user-perceived latency (i.e., the time between when the user speaks, and when words appear on the device). 
This included advancements such as end-to-end endpointing~\cite{Li2020} to encourage faster microphone closing; as well as FastEmit~\cite{Jiahui20} to encourage the model to emit tokens earlier.

Finally, in 2021 the model was further improved for the Pixel 6 smartphone~\cite{Sainath2021}, to take advantage of the tensor processing unit (TPU)~\cite{JouppiYoungPatilEtAl17} on the device. This includes using conformer layers for the encoder~\cite{Gulati+Qin+:2020conformer}; a small embedding prediction network for the decoder~\cite{Rami21}; a 2-pass cascaded encoder to run a 2nd-pass beam search~\cite{Arun21}; and, a neural LM re-scorer to help improve accuracy long-tail named entities. This model is the best ASR system that Google has released to date, both in terms of quality and latency, and is significantly better then the 80GB model in the cloud.

\section{Areas for Future Work}
\label{sec:future}
Currently, \EtoE\ models dominate the academic debate on ASR. This, at least partly, is not (yet?) reflected in the corresponding commercial deployment of \EtoE\ ASR architectures. \EtoE\ models are not yet the perfect match for all ASR conditions and further research is needed to take advantage of the benefits of \EtoE\ modeling. 

\EtoE\ models seem to perform really well when training data is abundant, while not scaling well to low-resource conditions. Similarly, domain change requires a flexible exchange of language models, which is natural for classical ASR models based on a separation of acoustic and language models. Ongoing research on the use of external language models in \EtoE\ models and internal language model estimation already is promising, but can be expected to see further improvements. 

Top \EtoE\ ASR systems usually require orders of magnitude more training epochs than comparable classical ASR systems, and further research into efficient and robust optimization and training schedules is needed. 

The high level of integration of \EtoE\ models also involves a loss in modularity, which might support the explainability and reusability of models. Also, more efficient training schedules might take advantage of modularity. One assumed advantage of \EtoE\ models is that everything is trained from data and secondary knowledge sources (e.g.\ pronunciation lexica and phoneme sets) are avoided. However, rare events, like rare words in ASR still provide a challenge, which needs further research.

With the missing separation of acoustic and language models, the question arises of how to exploit text-only resources in \EtoE\ model training - do we foresee solutions beyond training data generation using TTS?
We note that a number of recent works have explored approaches to combine speech and text modalities by attempting to implicitly or explicitly map them into a shared space~\cite{BapnaChungWuEtAl21, BapnaCherryZhangEtAl22, TangGongDongEtAl22, ChungZhuZeng21, AoWangZhouEtAl22, ThomasKuoKingsburySaon22, Chen+Zhang+:2022MAESTRO, SainathPrabhavalkarBapnaEtAl22}.
Furthermore, high-performance \EtoE\ solutions exist for both discriminative problems like ASR, as well as generative problems like TTS, how can both be exploited jointly to support semi-supervised training based on text-only and/or audio-only data on top of transcribed speech audio \cite{tjandra2017speechchain,Hori+:cycle-consistency:2018}?

More specifically, for the AED architecture we observe a length bias. Although many heuristics are known to tackle length bias in AED, we are still missing a well-founded explanation for it, as well as a  corresponding remedy of the original model.

Other open research problems include speaker adaptation and robustness to recording conditions, especially in mismatch situations.

The \EtoE\ principle also provides a promising candidate to solve multichannel ASR by providing an \EtoE\ solution jointly tackling the source separation, speaker diarization and speech recognition problem \cite{ochiai2017multichannel,chang2019mimo}.

Finally, we need to investigate, if \EtoE\ is a suitable guiding principle, and how different \EtoE\ ASR models relate to each other as well as to classical ASR approaches. The most important guiding principle of ASR research and development has been performance, and ASR has been boosted strongly by widely used benchmark tasks and international evaluation campaigns. With the current diversity of classical and \EtoE\ models, we also need to resolve the question of what constitutes state-of-the-art in ASR today, and can we expect a common state-of-the-art ASR architecture in the future?
          
\section{Conclusions}
In this work, we presented a detailed overview of end-to-end approaches to ASR.
Such models, which have grown in popularity over the last few years, propose to use highly integrated neural network components which allow input speech to be converted directly into output text sequences through character-based output units.
Thus, such models eschew the classical modular ASR architecture consisting of an acoustic model, a pronunciation model, and a language model, in favor of a single compact structure, and rely on the data to learn effectively. 
These design choices enable the deployment of highly accurate on-device speech recognition models (see Section~\ref{sec:applications}), but also come with a number of downsides which are still areas of active research (see Section~\ref{sec:future}).
Finally, we note that Jinyu Li's excellent overview article~\cite{li2022:e2ereview}, is complementary to our article, offering another perspective on end-to-end ASR.

\label{sec:conclusions}

\section*{Acknowledgment}
The authors would like to thank Julian Dierkes, Yifan Peng, Zolt\'an T\"uske, Albert Zeyer and Wei Zhou for their help on refining our manuscript.

\ifCLASSOPTIONcaptionsoff
  \newpage
\fi

\bibliographystyle{IEEEtran}
\bibliography{refs}

\begin{IEEEbiography}{Rohit Prabhavalkar}
Rohit Prabhavalkar received his PhD in Computer Science and Engineering from The Ohio State University, USA, in 2013. 
Following his PhD, Rohit joined the Speech Technologies group at Google where he is currently a Staff Research Scientist. 
At Google, his research has focused primarily on developing compact acoustic models which can run efficiently on mobile devices, and on developing improved end-to-end automatic speech recognition systems.
Rohit has co-authored over 50 refereed papers, which have received two best paper awards (ASRU 2017; ICASSP 2018).
He currently serves as a member of the IEEE Speech and Language Processing Technical Committee (2018--2024) and as an associate editor of the IEEE/ACM Transactions on Audio, Speech, and Language Processing.
\end{IEEEbiography}

\begin{IEEEbiography}{Takaaki Hori}
received his PhD degree in system and information engineering from Yamagata University, Yonezawa, Japan, in 1999. From 1999 to 2015, he had been engaged in researches on speech recognition and spoken language processing at Cyber Space Laboratories and Communication Science Laboratories in Nippon Telegraph and Telephone (NTT) Corporation, Japan. From 2015 to 2021, he was a Senior Principal Research Scientist at Mitsubishi Electric Research Laboratories (MERL), USA. He is currently a Machine Learning Researcher at Apple.
His research interests include automatic speech recognition, spoken language understanding, and language modeling. He served as a member of the IEEE Speech and Language Processing Technical Committee (2020--2022).
\end{IEEEbiography}

\begin{IEEEbiography}{Tara Sainath}
received her PhD in Electrical Engineering and Computer Science from MIT in 2009. The main focus of her PhD work was in acoustic modeling for noise robust speech recognition. After her PhD, she spent 5 years at the Speech and Language Algorithms group at IBM T.J. Watson Research Center, before joining Google Research. She has served as a Program Chair for ICLR in 2017 and 2018. Also, she has co-organized numerous special sessions and workshops, including Interspeech 2010, ICML 2013, Interspeech 2016 and ICML 2017. In addition, she is a member of the IEEE Speech and Language Processing Technical Committee (SLTC) as well as the Associate Editor for IEEE/ACM Transactions on Audio, Speech, and Language Processing.
\end{IEEEbiography}

\begin{IEEEbiography}{Ralf Schl\"{u}ter}
Ralf Schl\"{u}ter received his Dr.rer.nat.\ degree in Computer Science in 2000 and habilitated in Computer Science in 2019, both at RWTH Aachen University. In May 1996, Ralf Schl\"uter joined the Computer Science Department at RWTH Aachen University, where he currently is Lecturer and Academic Director, leading the Automatic Speech Recognition Group at the Chair Computer Science 6 -- Machine Learning and Human Language Technology. In 2019, Ralf also joined AppTek GmbH Aachen as Senior Researcher. His research interests cover sequence classification, specifically all aspects of automatic speech recognition, decision theory, stochastic modeling, and signal analysis. Ralf served as  Subject Editor for Speech Communication (2013-2019).
\end{IEEEbiography}

\begin{IEEEbiography}{Shinji Watanabe}
Shinji Watanabe is an Associate Research Professor at Johns Hopkins University, Baltimore, MD, USA. He received his PhD degree in 2006 from Waseda University, Tokyo, Japan. 
He was a research scientist at NTT Communication Science Laboratories, Kyoto, and a Senior Principal Research Scientist at Mitsubishi Electric Research Laboratories (MERL). 
His research interests include automatic speech recognition, speech enhancement, and machine learning for speech and language processing. 
He served an Associate Editor of the IEEE Transactions on Audio Speech and Language Processing, and is a member of several technical committees including the IEEE Signal Processing Society Speech and Language Technical Committee.
\end{IEEEbiography}

\end{document}